\documentclass{jfm}

\usepackage{afterpage}
\usepackage{longtable}
\usepackage[english]{babel}
\usepackage{dcolumn}
\usepackage{bm}
\usepackage{pifont}
\usepackage{amsfonts}
\usepackage{amsmath}
\usepackage{upmath}
\usepackage{amssymb}
\usepackage{wasysym}
\usepackage{amsbsy}
\usepackage{graphicx}
\usepackage{natbib}
\usepackage{psfrag}

\catcode`\@=11
\def\gtsim{\mathrel{\vcenter{\m@th\offinterlineskip
\hbox{$\hfill>\hfill$}\kern.5ex\hbox{$\hfill\sim\hfill$}}}}
\catcode`\@=12

\catcode`\@=11
\def\ltsim{\mathrel{\vcenter{\m@th\offinterlineskip
\hbox{$\hfill<\hfill$}\kern.5ex\hbox{$\hfill\sim\hfill$}}}}
\catcode`\@=12 \catcode`\@=12

\ifCUPmtlplainloaded \else
  \checkfont{eurm10}
  \iffontfound
    \IfFileExists{upmath.sty}
      {\typeout{^^JFound AMS Euler Roman fonts on the system,
                   using the 'upmath' package.^^J}%
       \usepackage{upmath}}
      {\typeout{^^JFound AMS Euler Roman fonts on the system, but you
                   dont seem to have the}%
       \typeout{'upmath' package installed. JFM.cls can take advantage
                 of these fonts,^^Jif you use 'upmath' package.^^J}%
      }
  \else
  \fi
\fi

\ifCUPmtlplainloaded \else
  \checkfont{msam10}
  \iffontfound
    \IfFileExists{amssymb.sty}
      {\typeout{^^JFound AMS Symbol fonts on the system, using the
                'amssymb' package.^^J}%
       \usepackage{amssymb}%
         \let\leq=\leqslant
         
      }{}
  \fi
\fi

\ifCUPmtlplainloaded \else
  \IfFileExists{amsbsy.sty}
    {\typeout{^^JFound the 'amsbsy' package on the system, using it.^^J}%
     \usepackage{amsbsy}}
    {\providecommand\boldsymbol[1]{\mbox{\boldmath $##1$}}}
\fi

\newcommand\Web{\mbox{\textit{We}}}  
\newcommand\Fr{\mbox{\textit{Fr}}}   
\newcommand\Oh{\mbox{\textit{Oh}}}   
\newcommand\Bo{\mbox{\textit{Bo}}}   
\newcommand\Mo{\mbox{\textit{Mo}}}   

\title[The effect of viscous relaxation on the spatiotemporal stability of capillary jets]
{The effect of viscous relaxation on the spatiotemporal stability of capillary jets}

\author[A. Sevilla]%
{Alejandro Sevilla} 

\affiliation{\'Area de Mec\'anica de Fluidos, Universidad Carlos III 
de Madrid, 28911 Legan\'es, Spain.}

\begin{document}
\maketitle

\begin{abstract}
The linear spatiotemporal stability properties of axisymmetric laminar capillary jets with 
fully developed initial velocity profiles are studied for large values of both the Reynolds 
number, $\Rey=Q/(\pi\,a\,\nu)$, and the Froude number, $\Fr=Q^2/(\pi^2\,g\,a^5)$, where $a$ 
is the injector radius, $Q$ the volume flow rate, $\nu$ its kinematic viscosity, and $g$ the 
gravitational acceleration. The downstream development of the basic flow and its stability 
are addressed with an approximate formulation that takes advantage of the jet slenderness. 
The base flow is seen to depend on two parameters, namely a Stokes number, 
$\text{G}=\Rey/\Fr$, and a Weber number, $\Web=\rho\,Q^2/(\pi^2\,\sigma\,a^3)$, where 
$\sigma$ is the surface tension coefficient, while its linear stability depends also on the 
Reynolds number. When non-parallel terms are retained in the local stability problem, the 
analysis predicts a critical value of the Weber number, $\Web_c(\text{G},\Rey)$, below which 
a pocket of local absolute instability exists within the near field of the jet. The function 
$\Web_c(\Rey)$ is computed for the buoyancy-free jet, showing marked differences with the 
results previously obtained with uniform velocity profiles. It is seen that, in accounting 
for gravity effects, it is more convenient to express the parametric dependence of the critical 
Weber number with use made of the Morton and Bond numbers, $\Mo=\nu^4 \rho^3 g/\sigma^3$ and 
$\Bo=\rho g a^2/\sigma$, as replacements for $\text{G}$ and $\Rey$. This alternative 
formulation is advantageous to describe jets of a given liquid for a known value of $g$, 
in that the resulting Morton number becomes constant, thereby leaving $\Bo$ as the only 
relevant parameter. The computed function $\Web_c(\Bo)$ for a water jet under Earth gravity 
is shown to be consistent with the experimental results of Clanet \& Lasheras 
(\emph{J. Fluid Mech.} vol. 383, 1999, p. 307) for the transition from jetting to dripping 
of water jets discharging into air from long injection needles, which cannot be properly 
described with a uniform velocity profile assumed at the jet exit.
\end{abstract}

\section{Introduction}\label{sec:intro}

The instability of capillary jets is a phenomenon of great importance in many applications 
like propulsion, irrigation, ink-jet printing and 
microfluidics~\cite[see the reviews][]{Bogy,Eggers1997,Lin1998,BarreroAR,EyV}.
One particular aspect of this flow which has received a lot of attention in the last decade, 
mainly due to its relevance in emerging microfluidic 
applications~\cite[see for instance][]{Basaran2002,UtadaMagic}, is the fact that a minimum 
flow rate is required to achieve a \emph{jetting regime}, in which droplets detach from the 
tip of a slender liquid column either due to the growth of capillary 
instabilities~\cite[][]{Rayleigh1878}, or an end-pinching 
mechanism~\cite[][]{GordilloTipBreakup}. In contrast, flow rates smaller than this minimum 
value give rise to a \emph{dripping regime}, in which drop formation takes place directly 
at the injector outlet, providing comparatively larger drops at a smaller 
frequency~\cite[][]{Wilkes1999,Ambravaneswaran2004}. In the simplest situation, in which 
the effect of the outer fluid can be neglected, the existence of these two regimes can be 
explained in physical terms by a competition between the inertia of the liquid, which 
promotes the formation of a slender jet, and the surface tension forces, which tend to 
destabilise it~\cite[see][]{EyV}. The relative importance of these forces is measured by 
a liquid Weber number, such that the dripping and jetting regimes prevail for values of the 
Weber number smaller and larger than certain critical values, respectively.\\

The fact that a slender capillary jet cannot be sustained for arbitrarily small values of 
the liquid Weber number was early recognised~\cite[see for instance][]{SmithMoss}. However, 
this \emph{global} destabilisation phenomenon cannot be explained within the classical 
temporal stability analysis of~\cite{Rayleigh1878}, and a theoretical explanation had to 
wait until the work of~\cite{LeibyGoldstein}, who found that a parallel capillary jet 
becomes absolutely unstable to inviscid axisymmetric perturbations for values of the Weber 
number below a certain critical value, which they found to depend on the specific shape of 
the axial velocity profile. Although these authors realised that ``the double root in the 
present flow will be associated with some natural droplet formation process'', they did 
not identify the onset of absolute instability in the jet as responsible for the transition 
from jetting to dripping. In the case of jets with uniform velocity profiles, 
\cite{LeibyGoldsteinAC} showed that the critical Weber number decreases as the Reynolds 
number decreases, indicating the stabilising effect of viscosity. The experiments
of~\cite{Monkewitz88cap} led to the conjecture that the jetting-dripping transition could 
be due to the destabilization of a linear global mode in the sense of~\cite{Chomaz1988}. 
This possibility was investigated by~\cite{ledizes97}, who performed a stability analysis 
of the falling capillary jet with uniform velocity profile for large Reynolds and Froude 
numbers. \cite{ledizes97} found that a sufficiently long region of absolute instability 
near the orifice does indeed destabilise a linear global mode due to the interaction of 
three spatial branches of the dispersion relation. The corresponding critical value of 
the Weber number, computed by~\cite{ledizes97} for Reynolds numbers between 100 and 200, 
was shown to be in good quantitative agreement with the experimental point 
of~\cite{Monkewitz88cap} for $\Rey\simeq 180$. It is interesting to note that the related 
configuration of a gas jet injected into stagnant liquid is absolutely unstable regardless 
of the value of the gas Weber number and, correspondingly, a bubbling regime is observed in 
experiments~\cite[][]{Oguz1993}. However, a convective instability type is also possible in 
this system when a sufficiently fast liquid coflow is imposed, as shown 
by~\cite{Sevilla2005a}, where the predictions of spatiotemporal stability theory were shown 
to be in good agreement with the experimental observation of the transition from a bubbling 
to a jetting regime.\\

A detailed experimental study of the jetting/dripping transition in water jets discharging 
from long needles was performed by~\cite{Clanet1999}, where a successful mechanical model 
for the dripping to jetting transition was also developed. In particular, the experiments 
of~\cite{Clanet1999} revealed the presence of hysteresis in the dripping/jetting and 
jetting/dripping transitions when the injector diameter is sufficiently large. In 
dimensionless terms, hysteresis appears at high enough values of the Bond number, for which 
the critical Weber number for the dripping to jetting transition is substantially larger 
than the corresponding value for the reverse process. Moreover, \cite{Clanet1999} found 
that the critical Weber number increases as the Bond number decreases, in qualitative, 
albeit not quantitative, agreement with the prediction of~\cite{ledizes97}. Moreover, the 
critical Weber number was found by~\cite{Clanet1999} to remain approximately constant for 
small enough values of the Bond number. Surprisingly, this limiting value of the critical 
Weber number is closer to the theoretical result for the appearance of absolute instability 
in an \emph{inviscid} jet with \emph{uniform} velocity profile \cite[][]{LeibyGoldstein}, 
than to the actual, expectedly more exact, prediction by~\cite{ledizes97}. A natural 
explanation for this discrepancy is that, while the theory of \cite{ledizes97} assumes a 
uniform velocity profile, the experiments of~\cite{Clanet1999} were performed with long 
needles for which the exit velocity profile is close to the fully-developed, parabolic 
Poiseuille profile. In fact, the analysis of~\cite{LeibyGoldstein} contemplated the 
influence of the velocity profile shape by introducing the parametric family 
$(1-br^2)/(1-b/2)$, where $r$ is the radial coordinate, and the parameter $b$ was varied 
between $0$ and $1$, respectively corresponding to uniform and parabolic profiles. The 
latter was found to be more stable, having a critical Weber number approximately seven times 
smaller than the uniform one. \cite{LeibyGoldstein} suggested that their parametric profiles 
could be used to model the viscous relaxation process of a capillary jet with parabolic 
initial velocity profile, with the downstream position corresponding to a certain value 
of the parameter $b$. However, the actual shapes of the velocity profiles within the 
relaxation region are different from the parametric family proposed 
by~\cite{LeibyGoldstein}. Moreover, the relationship between the downstream position and 
the value of the parameter $b$ is not uniquely defined.\\

The present work addresses the linear spatiotemporal stability properties of laminar 
capillary jets with fully-developed outlet velocity profiles, assuming large values of 
the Reynolds and Froude numbers in the analysis. In contrast with previous stability 
studies, which used parametric shapes to describe the downstream evolution of the jet 
velocity profiles, here the base flow is computed by integrating the axisymmetric boundary 
layer equations. The linear spatiotemporal stability of the resulting slender flow is then 
studied using a quasi-parallel approximation. The paper is organised as follows: first, in 
\S{\ref{sec:formulation}}, we present the model considered for the base flow, as well as 
for the computation of its linear stability. Then, in \S{\ref{sec:absolute}}, we present 
the results of the spatiotemporal stability analysis, and discuss its relationship with the 
phenomenon of transition to dripping observed in experiments. Finally, 
\S{\ref{sec:conclusions}} is devoted to the conclusions.

\section{Formulation of the problem}\label{sec:formulation}

\begin{figure}
\begin{center}
    \includegraphics[width=0.9\textwidth]{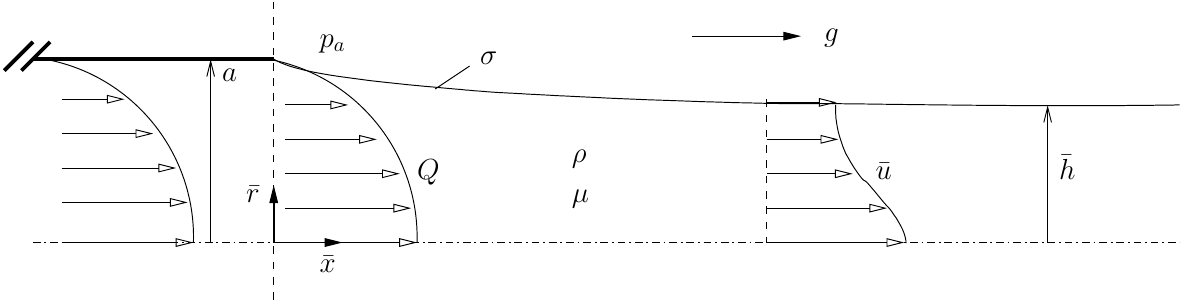}
    \caption{\label{fig:scheme}Sketch of the flow configuration considered in the present 
work at a given instant.}
\end{center}
\end{figure}

The configuration under study consists of a flow rate $Q$ of liquid with density $\rho$ 
and viscosity $\mu$ discharging downwards from a tube of radius $a$ and length $l_t$ into 
a stagnant atmosphere of gas of density $\rho_g$ at pressure $p_a$. The tube wall is assumed 
vanishingly thin, so that $a$ and $l_t$ are the only relevant geometric length scales for 
the jet. Note that this formulation would also apply to tubes of finite thickness if the 
liquid does not wet the exit section, so that the contact line is pinned at the inner edge 
of the tube at its outlet. To describe the resulting liquid jet we will adopt a cylindrical 
coordinate system with its origin placed at the center of the exit section and the axial 
coordinate pointing in the direction of gravity, as depicted in figure~\ref{fig:scheme}. 
We shall assume values the Reynolds number such that $1\ll\Rey=Q/(\pi\,a\,\nu)\lesssim 1000$, 
the latter limit in order to ensure laminar flow in the jet, as well as values of the Froude 
number $\Fr=Q^2/(\pi^2\,g\,a^5)\gg 1$, where $g$ is the gravitational acceleration. Keeping 
in mind that jet formation requires values of the Weber number 
$\Web=\rho\,Q^2/(\pi^2\,\sigma\,a^3)\gtrsim\mathcal{O}(1)$, and that $\Rey=\Web^{1/2}/\Oh$, 
the first condition is fulfilled whenever the Ohnesorge number 
$\Oh=\mu/(\rho\,\sigma\,a)^{1/2}\ll 1$, where $\sigma$ is the surface tension coefficient. 
Notice that $\Oh$ does not depend on the jet velocity, and can be seen as a Reynolds number 
based on the capillary velocity, $(\sigma/(\rho\,a))^{1/2}$.\\

With these assumptions, the free laminar liquid jet which appears downstream of the tube 
exit contracts due to the action of gravity, as well as to the viscous relaxation of the 
initial velocity profile, which is non-uniform at the injector exit due to the accumulated 
action of the viscous shear stress at the inner wall. The exact shape of this velocity 
profile depends on the detailed geometry of the injection system as well as on the 
Reynolds number, precluding a universal description of the resulting free jet. In this 
work we will consider the limiting case of a jet discharging from a long cylindrical tube 
with $l_t\gtrsim \Rey\,a\gg a$, such that the exit velocity profile is the fully-developed 
Poiseuille profile, $\bar{u}_0=2U_0(1-\bar{r}^2/a^2)$, where $U_0=Q/\pi\,a^2$ is the mean 
exit velocity. Throughout the text, quantities with bar will denote dimensional variables. 
Laminar liquid jets with fully developed velocity profiles at the outlet appear frequently 
in applications. In particular, this is the case of the experiments of~\cite{Clanet1999}, 
devoted to the study of transition from dripping to jetting in water jets under Earth gravity. 
However, it is important to emphasise that the hypodermic needles used by~\cite{Clanet1999} 
had finite wall thickness, and the liquid wetted their entire exit section. Therefore, in 
those experiments the contact line was pinned at the \emph{outer} edge and, consequently, 
the thin wall configuration considered in the present work cannot be considered fully 
equivalent to the flow experimentally studied by~\cite{Clanet1999}. In particular, in 
contrast with the simplified boundary layer description used here, flow separation and 
recirculation are expected to take place in a small region near the needle exit section 
whenever the contact line is pinned at the outer edge. The structure of the flow in this 
region must be studied with the full Navier-Stokes equations, and is outside the scope of 
the present investigation. Keeping in mind the difference between both configurations, and 
in the absence of experimental data corresponding to the thin wall limit, the results 
of~\cite{Clanet1999} will be compared in~\S{\ref{subsec:water}} with those obtained in the 
present work.\\

As mentioned above, the streamwise contraction of the jet is due to a combined contribution 
of gravity and viscous relaxation. Concerning the former, the characteristic axial distance 
$l_g$ over which gravity accelerates the jet by a quantity of the order of its initial velocity 
is given by a balance between the convective acceleration of the liquid, 
$\bar{u}\,\partial\bar{u}/\partial\bar{x}\sim U_0^2/l_g$, and the gravitational acceleration, 
$g$, so that $l_g\sim\Fr\,a$. On the other hand, viscous relaxation takes place within a 
distance $l_r$ of the exit such that the convective acceleration $U_0^2/l_r$ balances the 
viscous diffusion term, of order $\nu\,U_0/a^2$, giving $l_r\sim\Rey\,a$. From these 
considerations it is clear that the conditions $\Fr\gg 1,\,\Rey\gg 1$ suffice to ensure 
the slenderness of the jet, whose downstream evolution will be described here by means of 
the following decomposition,
\begin{eqnarray}
 \bar{u} &=& U_0\left[U(X,r) + \epsilon u'(x,r,t)\right],\label{eq:decu}\\
 \bar{v} &=& U_0\left[\Rey^{-1}\,V(X,r) + \epsilon v'(x,r,t)\right],\label{eq:decv}\\
 \bar{p} &=& \rho U_0^2\left[P(X,r) + \epsilon p'(x,r,t)\right],\label{eq:decp}\\
 \bar{h} &=& a\left[H(X) + \epsilon h'(x,t)\right],\label{eq:dech}
\end{eqnarray}
where $x=\bar{x}/a$, $r=\bar{r}/a$, $t=\bar{t}\,U_0/a$. Here, $\left(U,V,P,H\right)$ represent 
the steady basic flow, which is non-dimensionalised following the boundary-layer scaling, for 
which $X=x/\Rey$ and $\bar{U}\sim\Rey\,\bar{V}\sim U_0$, and $\epsilon$ is a small parameter 
measuring the size of the unsteady perturbations around the mean flow, $\left(u',v',p',h'\right)$.

\subsection{The basic flow}\label{subsec:basicflow}

The equations governing the basic flow are obtained by substituting the decomposition given by 
equations (\ref{eq:decu})-(\ref{eq:dech}) into the incompressible, axisymmetric Navier-Stokes 
equations to yield, at $\mathcal{O}(\epsilon^0,\,\Rey^{-1})$, the steady axisymmetric 
boundary-layer equations,
\begin{eqnarray}
U_X+\textstyle{\frac{1}{r}}(rV)_r&=&0,\label{eq:bl1}\\
U\,U_X+V\,U_r&=&\Web^{-1}H^{-2}H_X+
\textstyle{\frac{1}{r}}\left(r\,U_r\right)_r+\text{G},\label{eq:bl2}
\end{eqnarray}
where $\text{G}=\Rey/\Fr$ is a Stokes number, and subscripts denote partial derivatives. Notice 
that both the axial viscous stress and the radial pressure gradient have been neglected, since 
they are $\mathcal{O}(\Rey^{-2})\ll 1$. Equations (\ref{eq:bl1})-(\ref{eq:bl2}) are supplemented 
with the initial condition $X=0:\,U=u_0(r)=2(1-r^2)$, the boundary conditions $r=0:\,V=U_r=0$ 
and $r=H:\,U_r=0$, and the mass conservation constraint $\int_0^H Urdr=1/2$. In the boundary 
condition at the free surface $r=H$, we have neglected terms $\mathcal{O}(\Rey^{-2})\ll 1$ 
or smaller in the continuity of tangential stress at the interface. Moreover, we have also 
made the assumption that the external gas, of density $\rho_g$, has a negligible influence 
in the dynamics, which is a good approximation since the aerodynamic Weber number, 
$\Web_g=\Web\,\rho_g/\rho\ll 1$ \cite[see][]{Weber1931,GordilloFirstWind}. In effect, note 
that $\rho_g/\rho\ll 1$, and the critical values of the Weber number $\Web_c\sim\mathcal{O}(1)$, 
as will be shown below. Finally, the axial pressure gradient $-P_X$ was obtained by the normal 
stress balance at the interface, where we have neglected both the axial curvature, with a 
relative error $\text{max}[\mathcal{O}(\Fr^{-2}),\,\mathcal{O}(\Rey^{-2})]\ll 1$, and the 
normal viscous stress at the interface, with an error $\mathcal{O}(\Rey^{-2})\ll 1$. Thus, 
the pressure gradient is given by $-P_X=\Web^{-1}H^{-2}H_X$ with a small error 
$\text{max}[\mathcal{O}(\Fr^{-2}),\,\mathcal{O}(\Rey^{-2})]\ll 1$.
\begin{figure}
\begin{center}
    \includegraphics[width=0.9\textwidth]{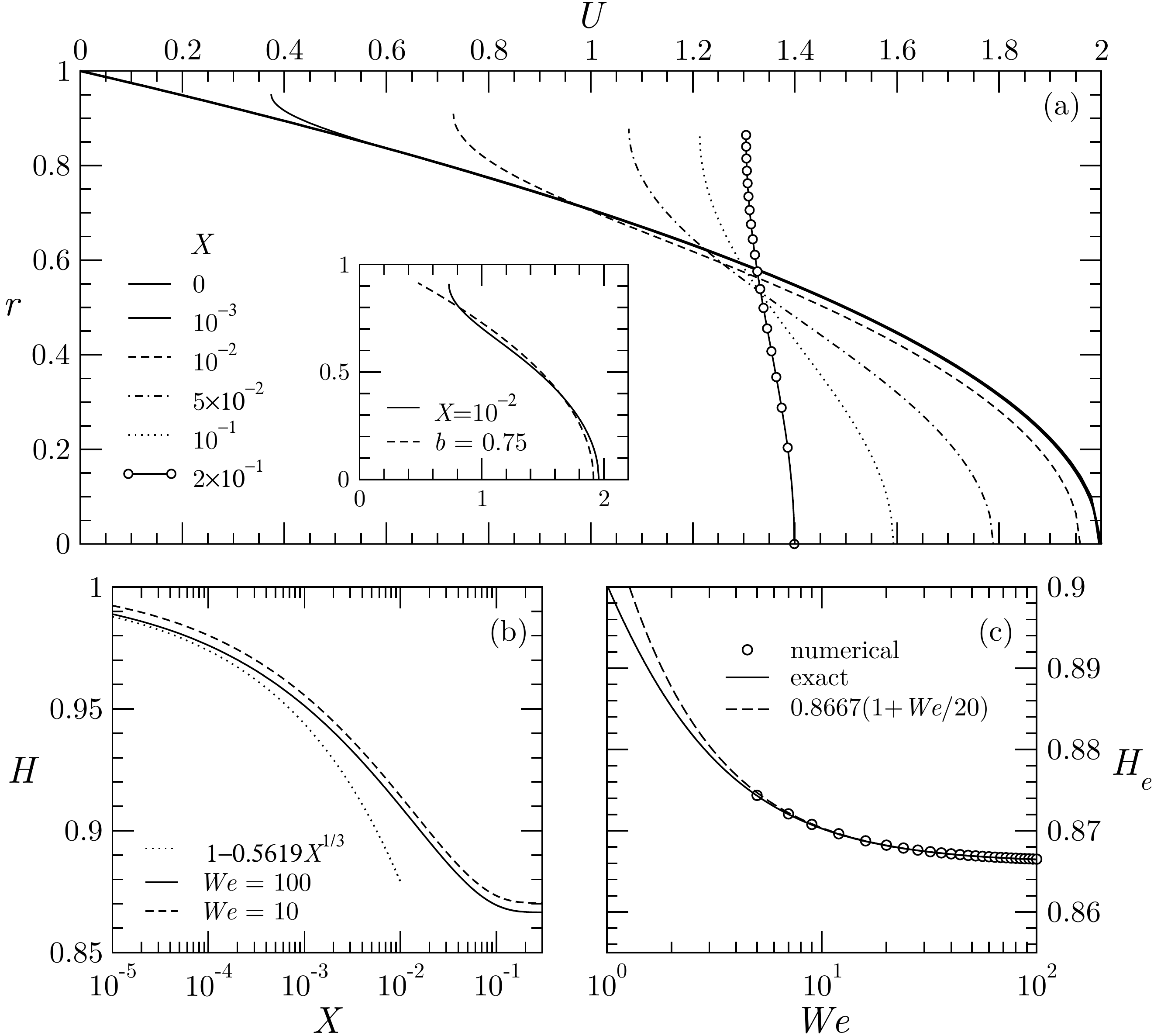}
    \caption{\label{fig:baseflowg0} The basic flow for $\text{G}=0$. (a) Velocity profiles 
for $\Web=10$ at several downstream positions $X$, indicated in the legend. The inset shows 
a comparison of the velocity profile at $X=0.01$, with a profile taken from the parametric
family of~\cite{LeibyGoldstein} for a value of the parameter $b=0.75$. (b) Downstream 
evolution of the jet radius for $\Web=10$ (solid line) and $\Web=100$ (dashed line), together 
with the similarity solution due to~\cite{Goren66} $H=1-0.5619\,X^{1/3}$ (dotted line), and 
(c) asymptotic jet radius $H_e$ as a function of $\Web$ computed numerically (circles), 
compared with the exact solution (solid line) and its large Weber number approximation 
$\sqrt{3}/2\,(1+\Web/20)$ (dashed line).}
\end{center}
\end{figure}
An efficient and accurate method of lines was implemented to obtain the numerical solution 
of equations~(\ref{eq:bl1})-(\ref{eq:bl2}), as explained in detail by~\cite{Gordillo2001}. 
Typically, 200 lines were used in the computations.\\

The boundary layer equations (\ref{eq:bl1})-(\ref{eq:bl2}), complemented with the boundary 
conditions discussed in the previous paragraph, have been studied in detail in the past. 
Numerical solutions in several representative cases were given by~\cite{DudayVrentas}, 
while~\cite{OguzRelaxation} provided a semi-analytical approach based on eigenfunction 
expansions. In addition, these studies showed good agreement between experiments and 
equations~(\ref{eq:bl1})-(\ref{eq:bl2}). The very near field of the jet, $X\to 0$, was 
considered by~\cite{Goren66}, who showed that the structure of the viscous layer which 
develops at the free surface resembles a Goldstein wake, while~\cite{Tillett68} extended 
the analysis to include higher-order effects. On the other hand, the far field of the jet 
was considered by~\cite{GorenyWronski}, who provided with a complete description of the 
approach of the jet to the uniform conditions achieved for $X\gg 1$. Finally, asymptotic 
solutions uniformly valid throughout the entire jet were given by~\cite{franceses}.

The main features of the basic flow considered here can be seen in 
figure~\ref{fig:baseflowg0}, computed for the particular case of negligible gravitational 
effects, $\text{G}=0$. Several examples of numerically computed velocity profiles along 
the relaxation region for $\Web=10$ and different values of $X$ are shown in
figure~\ref{fig:baseflowg0}(a). Notice that the initial region of jet acceleration is 
confined to a thin viscous layer adjacent to the free surface, as revealed by the velocity 
profile represented with a thin solid line for $X=10^{-3}$. Mass conservation implies that 
the jet core must decelerate for larger values of $X$, as revealed by the velocity profile 
at $X=10^{-2}$ shown as the thin dashed line in figure~\ref{fig:baseflowg0}(a). The 
acceleration at the free surface, together with the deceleration at the axis, finally lead 
to an almost uniform velocity profile at $X=0.2$, shown with circles in 
figure~\ref{fig:baseflowg0}(a). A common feature of the local velocity profiles during 
relaxation is that they have an inflection point, whose radial position moves from the 
free surface towards the center of the jet as $X$ increases. Note, however, that the 
velocity profiles proposed by~\cite{LeibyGoldstein} to model the relaxation region, namely 
$U_{\text{LG}}=(1-br^2)/(1-b/2)$ with $0\leq b\leq 1$, are substantially different from the 
actual profiles in the relaxation region. This fact is illustrated by the inset of 
figure~\ref{fig:baseflowg0}(a), which shows a comparison between the velocity profile 
obtained through equations~(\ref{eq:bl1})-(\ref{eq:bl2}) at $X=0.01$ (solid line), and a 
profile taken from the family proposed by~\cite{LeibyGoldstein}, for a value of $b=0.75$ 
(dashed line). Notice that, to perform the comparison, the parametric profile has been 
properly renormalised according to the expression 
$\tilde{U}_{\text{LG}}(r)=[1-b(\tilde{r}/H)^2]/[H^2(1-b/2)]$, with $0\leq \tilde{r}\leq H$, 
such that $\int_0^H \tilde{r}U(\tilde{r})d\tilde{r}=1/2$. The downstream evolution of the 
unperturbed jet radius, $H(X)$, is shown in~\ref{fig:baseflowg0}(b) for two different values 
of the Weber number, namely $\Web=100$ (solid line) and $\Web=10$ (dashed line). Also shown 
in figure~\ref{fig:baseflowg0}(b) is the similarity solution $H=1-0.5619 X^{1/3}$, obtained
by~\cite{Goren66} for $X\to 0$ and $\Web\to\infty$, which is seen to be in good agreement 
with the numerical solution for $\Web=100$ and $X\lesssim 10^{-4}$. Another interesting 
feature of the jet, already pointed out by~\cite{OguzRelaxation}, is the fact that the 
radius relaxes sooner than the velocity profile. In effect, note from 
figure~\ref{fig:baseflowg0}(b) that $H$ takes its asymptotic value for $X\simeq 0.1$. In 
contrast, the dotted line represented in~\ref{fig:baseflowg0}(b) reveals that the velocity 
profile is still far from uniform at this station, the velocities being 
$U(X=0.1,r=0)\simeq 1.6$ and $U(X=0.1,r=H)\simeq 1.2$ at the centerline and the free 
surface, respectively. When surface tension effects are negligible, $\Web\gg 1$, it was 
shown by \cite{Harmon1955} the jet velocity and radius reach limiting values $U_e=4/3$ 
and $H_e=\sqrt{3}/2\simeq 0.866$ at distances $X\sim\mathcal{O}(1)$ downstream, as can be 
shown by performing a global mass and momentum balance for the jet. More generally, 
multiplying equation (\ref{eq:bl2}) by $r$, integrating across the jet, and making use of 
equation (\ref{eq:bl1}), one obtains the following equation for the downstream evolution 
of the axial momentum of the jet, $\text{J}(X)=\int_0^H U^2rdr$,
\begin{equation}
 \dfrac{d\text{J}}{dX}=\dfrac{1}{2\Web}\dfrac{dH}{dX}+\text{G}\dfrac{H^2}{2},\label{eq:J}
\end{equation}
which shows that, since $dH/dX<0$, surface tension removes momentum from the jet, while 
gravity adds to it. In the particular limit $G\ll 1$, the last term of equation (\ref{eq:J}) 
is negligible in the region $0\leq X\lesssim \mathcal{O}(1)$ where viscous relaxation takes 
place, and, integrating from $X=0$ to $X=X_e\sim\mathcal{O}(1)$, where $X_e$ is a station 
where the jet velocity profile is uniform, we obtain $U_e^2H_e^2/2-\text{J}_0=(H_e-1)/(2\Web)$. 
In addition, mass conservation implies that $U_eH_e^2=1$, and thus the relaxed jet radius for 
$G\ll 1$ is obtained from the cubic $H_e^3+(2\text{J}_0\Web-1)H_e^2-\Web=0$, first obtained 
by~\cite{Gavis1964} for the particular case of a parabolic velocity profile at the jet exit, 
and whose solution has the expansion
$H_e=(2\text{J}_0)^{-1/2}\left[1+\left(1-(2\text{J}_0)^{-1/2}\right)/(4\text{J}_0\Web)
+\mathcal{O}(\Web^{-2})\right]$ for $\Web\gg 1$. The initial momentum flux,
$\text{J}_0=\int_0^1U_0^2rdr$, has the value $\text{J}_0=2/3$ for the parabolic profile and, 
in this case, 
$H_e=\sqrt{3}/2[1+3(2-\sqrt{3})/(16\Web)+\mathcal{O}(\Web^{-2})]\approx\sqrt{3}/2(1+0.05/\Web)$, 
showing that the contribution of $\Web$ rapidly becomes negligible as $\Web\gtrsim 1$. The 
function $H_e(\Web)$ is displayed in figure~\ref{fig:baseflowg0}(c), where the numerically 
computed values of $H_e$, represented as open circles, are shown to be in very good agreement 
with the exact value, which solves the equation $H_e^3+(4\Web/3-1)H_e^2-\Web=0$, and which is
represented as a solid line in figure~\ref{fig:baseflowg0}(c). Finally, the fact that the 
numerical results accomplish the global mass and momentum balance to a very high degree of 
accuracy, together with the good agreement found with Goren's similarity solution for 
$X\to 0$, serve as a stringent validation for the numerical code implemented here for the 
base flow.

\subsection{The linear stability problem}\label{subsec:stability}

The equations governing the linear dynamics of axisymmetric disturbances to the basic jet 
considered in~\S{\ref{subsec:basicflow}} are obtained at $\mathcal{O}(\epsilon,\Rey^{-1})$ 
when~(\ref{eq:decu})-(\ref{eq:dech}) are substituted in the incompressible Navier-Stokes 
equations, leading to the following system of linearised stability equations,
\begin{eqnarray}
u'_x+\textstyle{\frac{1}{r}}(rv')_r&=&0,\label{eq:stab1}\\
u'_t+U\,u'_x+U_r\,v'+p'_x-\Rey^{-1}\left[\textstyle{\frac{1}{r}}\left(ru'_r\right)_r+
u'_{xx}\right]&=&
-\Rey^{-1}\left(U_X\,u'+V\,u'_r\right),\label{eq:stab2}\\
v'_t+U\,v'_x+p'_r-\Rey^{-1}\left[\left(\textstyle{\frac{1}{r}}\left(rv'\right)_r\right)_r+
v'_{xx}\right]&=&
-\Rey^{-1}\left(V_r\,v'+V\,v'_r\right),\label{eq:stab3}
\end{eqnarray}
to be solved with symmetry boundary conditions at the axis, $r=0: v'=u'_r=p'_r=0$, 
as well as the following linearised boundary conditions at the free surface, $r=H(X)$:
\begin{eqnarray}
&&v'-h'_t-U\,h'_x=\Rey^{-1}\left[H_X\,u'-V_r\,h'\right],\label{eq:sbc2}\\
&&p'-2\,\Rey^{-1}\,v'_r+\Web^{-1}\left(H^{-2}h'+h'_{xx}\right)=
-\left(\Web\Rey\right)^{-1}H^{-1}H_X\,h'_x,\label{eq:sbc3}\\
&&u'_r+v'_x+U_{rr}\,h'=
2\,\Rey^{-1}\left[\left(u'_x-v'_r\right)H_X+\left(U_X-V_r\right)h'_x\right],\label{eq:sbc4}
\end{eqnarray}
where equations~(\ref{eq:sbc2})-(\ref{eq:sbc4}) represent, respectively, the kinematic 
condition, the normal stress balance at the interface and the condition of zero shear stress 
at the interface. Notice that, as discussed in~\S\ref{subsec:basicflow}, the effect of the 
ambient gas has been neglected in equations~(\ref{eq:sbc3}) and~(\ref{eq:sbc4}). The terms 
appearing in the right hand sides of equations~(\ref{eq:stab1})-(\ref{eq:sbc4}) represent 
non-parallel effects, and have been included since, as will be shown below, they have a 
crucial effect on the the stability of the jet in the initial relaxation region.\\

Taking into account the fact that the most unstable perturbations to the basic jet have 
wavenumbers of the order of the jet radius, and thus evolve on the fast scale $x$, while 
the unperturbed jet evolves on the slow scale $X=x/\Rey$, a quasi-parallel approach will 
be used to solve the system of equations~(\ref{eq:stab1})-(\ref{eq:sbc4}), whereby 
axisymmetric disturbances are represented as normal modes of the form
\begin{equation}
\left(u',v',p',h'\right)=\left[\hat{u}(r),\hat{v}(r),\hat{p}(r),\hat{h}\right]
e^{i\left(k x-\omega t\right)},\label{eq:wkb}
\end{equation}
where quantities with hats are the complex perturbation amplitudes, and the phase function 
incorporates the evolution of the local instability wave on the fast scale $x$. The 
substitution of equation~(\ref{eq:wkb}) into the system~(\ref{eq:stab1})-(\ref{eq:sbc4}) 
leads to the following system of linear equations for the normal modes,
\begin{eqnarray}
ik\hat{u}+\textstyle{\frac{1}{r}}(r\hat{v})_r &=& 0,\label{eq:pse1}\\
i(kU-\omega)\hat{u}+U_r\,\hat{v}+ik\hat{p}-
\Rey^{-1}\left[\textstyle{\frac{1}{r}}\left(r\hat{u}_r\right)_r-k^2\hat{u}\right] 
&=& -\Rey^{-1}\left(U_X\,\hat{u}+V\,\hat{u}_r\right),\label{eq:pse2}\\
i(kU-\omega)\hat{v}+\hat{p}_r-\Rey^{-1}\left[\left(\textstyle{\frac{1}{r}}
\left(r\hat{v}\right)_r\right)_r-k^2\hat{v}\right] 
&=& -\Rey^{-1}\left(V_r\,\hat{v}+V\,\hat{v}_r\right),\label{eq:pse3}
\end{eqnarray}
which must be solved subjected to the symmetry conditions at the axis, 
$r=0: \hat{v}=\hat{u}_r=\hat{p}_r=0$, as well as boundary conditions at $r=H(X)$ 
resulting from the transformation of equations~(\ref{eq:sbc2})-(\ref{eq:sbc4}) 
through equation~(\ref{eq:wkb}),
\begin{eqnarray}
&&\hat{v}-i(kU-\omega)\hat{h} = \Rey^{-1}\left(H_X\,\hat{u}-V_r\,\hat{h}\right),\label{eq:bcpse1}\\
&&\hat{p}-2\,\Rey^{-1}\,\hat{v}_r+\Web^{-1}\left(H^{-2}-k^2\right)\hat{h} = 
-(\Web\Rey)^{-1}H^{-1}\,H_X\,ik\hat{h},\label{eq:bcpse2}\\
&&\hat{u}_r+ik\hat{v}+U_{rr}\,\hat{h} = 2\,\Rey^{-1}\left[\left(ik\hat{u}-\hat{v}_r\right)H_X 
+ \left(U_X-V_r\right)ik\hat{h}\right].\label{eq:bcpse3}
\end{eqnarray}
Since in this work we focus on the critical conditions for jet formability, we have solved 
the local eigenvalue problem, given by equations~(\ref{eq:pse1})-(\ref{eq:bcpse3}), with the 
aim at determining the convective or absolute nature of the local instability as a function
of the downstream position $X$. The basic hypothesis underlying the present approach is that 
the \emph{global} stability properties of the slender unperturbed jet obtained 
in~\S\ref{subsec:basicflow} can be inferred from the \emph{local} convective or absolute
nature of the instability as a function of $X$. In this context, two different scenarios 
have been clearly identified in the particular case of weakly non-parallel 
flows~\citep[for a review, see][]{Chomaz2005}: if the region of absolute instability (AI) 
starts right at the upstream boundary of the domain, the extent of the AI region must be 
larger than a certain minimum value for the global mode to be 
destabilised~\citep{Couairon1999}. In contrast, if the AI region is localised away from 
boundaries, a global mode is excited independently of its size~\citep{Pier98}. Although 
these findings have been rigorously demonstrated only for one-dimensional Ginzburg-Landau 
models, their application to actual fluid flows has been remarkably 
successful~\citep[see for instance][]{Pier2001,Lesshafft2006}.

Notice that, although the non-parallel terms in the right hand sides of 
equations~(\ref{eq:pse1})-(\ref{eq:bcpse3}) are usually disregarded when performing 
quasi-parallel stability analyses \citep[however, see e.g.][]{GordilloFirstWind,Herrada2008}, 
they are formally of the same order as the viscous terms, namely $\mathcal{O}(\Rey^{-1})$, 
and cannot be neglected except if $U_X\ll 1$ and $V\ll 1$. The latter conditions are not 
fulfilled in the initial relaxation region and, consequently, these non-parallel terms can 
be anticipated to play an important role in the local instability of the jet's near field. 
It is also important to emphasise that the quasi-parallel approach adopted in the present 
work is not completely rigorous, since base flow non-parallelism is accounted for only 
partially, through the right hand sides of equations~(\ref{eq:pse1})-(\ref{eq:bcpse3}). 
Alternative approaches to study the stability of the jet, like the parabolised stability 
equations~\citep{Herbert1997}, the linear global mode theory for weakly non-parallel 
flows~\citep{Chomaz1988,Huerre1990,monk93,ledizes97}, or a fully two-dimensional global 
stability analysis~\citep{Theofilis2011}, though more rigorous, are also substantially 
more involved than the quasi-parallel approach; they certainly constitute interesting 
topics for future work on the problem, but they are out of the scope of the present study.

The local stability problem given by the system of equations (\ref{eq:pse1})-(\ref{eq:bcpse3}) 
was solved with a very efficient Chebyshev collocation method. For each value of $X$, the 
radial coordinate $r$ was transformed into the basic interval $-1\leq\xi\leq 1$ by means of 
the linear transformation $r=(1-\xi)H(X)/2$. Introducing the vector of unknowns 
$\phi=(\hat{u}_i,\hat{v}_i,\hat{p}_i,\hat{h},k\hat{u}_i,k\hat{v}_i,k\hat{h})^T,\,i=1,\ldots,N$,
equations (\ref{eq:pse1})-(\ref{eq:bcpse3}) can be written as a generalised eigenvalue problem 
of the form $A\phi=kB\phi$, where $A$ and $B$ are complex matrices of size $(5N+2)\times(5N+2)$. 
The derivatives are computed in physical space at the set of $N$ Chebyshev collocation points, 
and the boundary conditions were implemented in the rows of the differentiation matrix 
corresponding to the points $r=0$ and $r=H$. A value of $N=30$ was enough to achieve very 
precise results.

\section{Absolute instability in the jet: transition from jetting to dripping}
\label{sec:absolute}

This section is devoted to the presentation and analysis of the results obtained for the 
spatiotemporal stability of the jet.

\subsection{Results for $G=0$}\label{subsec:G0}

Let us first consider the limit of negligible gravitational effects, $G\to 0$, in which 
case $\Web$ and $\Rey$ are the only parameters of the problem. Correspondingly, the 
critical Weber number for the onset of absolute instability is a function $\Web_c(\Rey)$,
which was obtained by \cite{LeibyGoldsteinAC} in the particular case of a jet of constant 
radius and uniform velocity profile.
\begin{figure}
\begin{center}
    \includegraphics[width=0.9\textwidth]{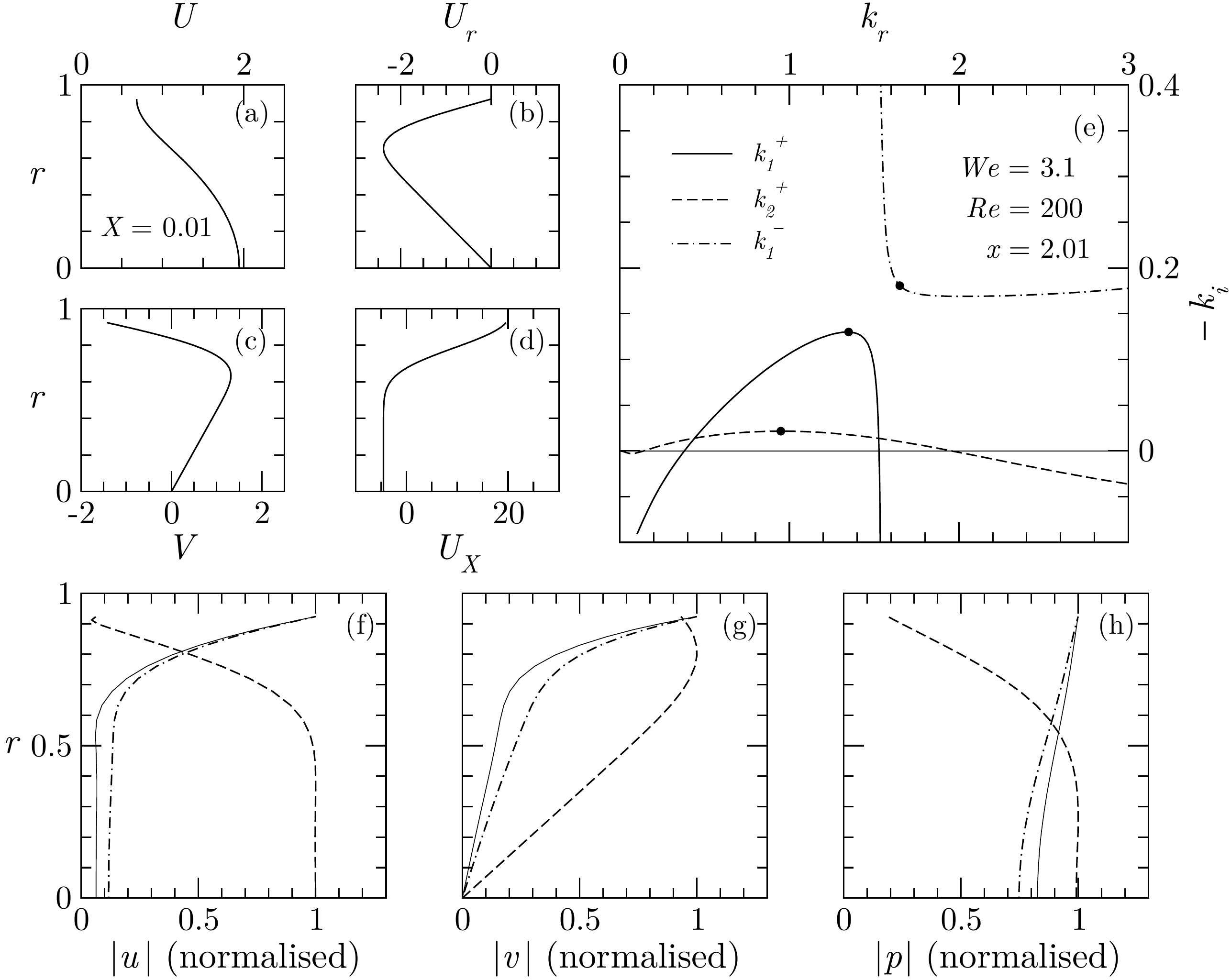}
    \caption{\label{fig:modes} Radial profiles of (a) $U$, (b) $U_r$, (c) $V$ and (d) $U_X$, 
for $\Web=3.1$ and $X=0.01$. (e) Corresponding spatial stability branches for $\Rey=200$. 
Normalised moduli of the eigenfunctions corresponding to the eigenvalues of the three spatial 
modes shown as solid circles in panel (e): (f) axial velocity, $|u|$, (g) radial velocity, 
$|v|$ and (h) pressure, $|p|$.}
\end{center}
\end{figure}
Figure \ref{fig:modes} summarises the results obtained for a jet with $\Web=3.1$ at the 
particular downstream position $X=0.01$. Panels~\ref{fig:modes}(a)-\ref{fig:modes}(d) 
show the corresponding base flow profiles of $U,U_r,V$ and $U_X$, respectively, which 
appear as coefficients in the stability equations (\ref{eq:pse1})-(\ref{eq:bcpse3}) and 
in their boundary conditions. Note, in particular, that the non-parallel convective terms 
$\Rey^{-1}\left(U_X\,\hat{u}+V\,\hat{u}_r\right)$ and 
$\Rey^{-1}\left(V_r\,\hat{v}+V\,\hat{v}_r\right)$ respectively appearing in equations 
(\ref{eq:pse2}) and (\ref{eq:pse3}) will be important for $X\ll 1$ and moderately large 
values of $\Rey$. Figure \ref{fig:modes}(e) shows three spatial stability branches 
corresponding to this particular basic flow for $\Rey=200$: two of them, denoted by 
$k_1^+$ (solid line) and $k_2^+$ (dashed line), contribute to the downstream response 
of the jet to forcing, while the $k_1^-$ branch (dash-dotted line) contributes to the 
upstream response. It must be emphasised that many other \emph{stable} spatial branches 
were found, but only these three are relevant to the present work. The fact that the 
two branches $k_1^+$ and $k_2^+$ contribute to the downstream response can be deduced 
from their movement in the $k$-plane when $\omega_i$ is varied: for values of $\omega_i$ 
larger than the maximum temporal growth rate for the parameter combination corresponding 
to figure \ref{fig:modes}, these two branches belong to the $k_i>0$ half-plane, whereas 
the branch $k_1^-$ remains in the opposite one, with $k_i<0$. Notice also from figure 
\ref{fig:modes}(e) that both downstream branches have unstable frequency ranges in which 
$k_i<0$, corresponding to linear waves with exponentially increasing amplitudes downstream 
from their source. On the contrary, branch $k_1^-$ is stable, leading to exponentially 
decreasing amplitudes upstream from the source. To provide a more complete picture of 
these spatial instability modes, figures \ref{fig:modes}(f)-(h) respectively show the 
normalised moduli of the eigenfunctions of axial velocity, $|u|$, radial velocity, $|v|$ 
and pressure, $|p|$, for the eigenvalues marked with a solid circle on top of the 
corresponding spatial branch in figure \ref{fig:modes}(e). Figures \ref{fig:modes}(f)-(h)
reveal that the eigenfunctions of the spatial modes $k_1^+$ (solid lines) and $k_1^-$
(dash-dotted lines), reach their maxima at the interface, $r=H$, and in fact resemble the 
Rayleigh modes appearing in unstretched jets with uniform velocity profiles. It is also 
interesting to note that non-parallel effects break the translation invariance, introducing 
a low wavenumber cut-off to the unstable $k_1^+$ branch, such that $\omega_i<0$ for $k_r=0$. 
On the other hand, the eigenfunctions associated to the unstable downstream branch $k_2^+$, 
shown as dashed lines in figures \ref{fig:modes}(f)-(h), indicate that this 
mode is of a different nature, its maximum amplitude taking place inside the jet. A 
parametric study, not shown here for brevity of presentation, reveals that the $k_2^+$ mode 
is unstable only for large enough values of both $\Rey$ and $\Web$, and small enough values 
of $X$. Thus, it is a \emph{surface wave mode}, which arises due to the interaction of shear 
and interface deflection \cite[see][for a detailed account]{Miles1960,SmithDavis1982}.\\

\begin{figure}
\begin{center}
    \includegraphics[width=0.9\textwidth]{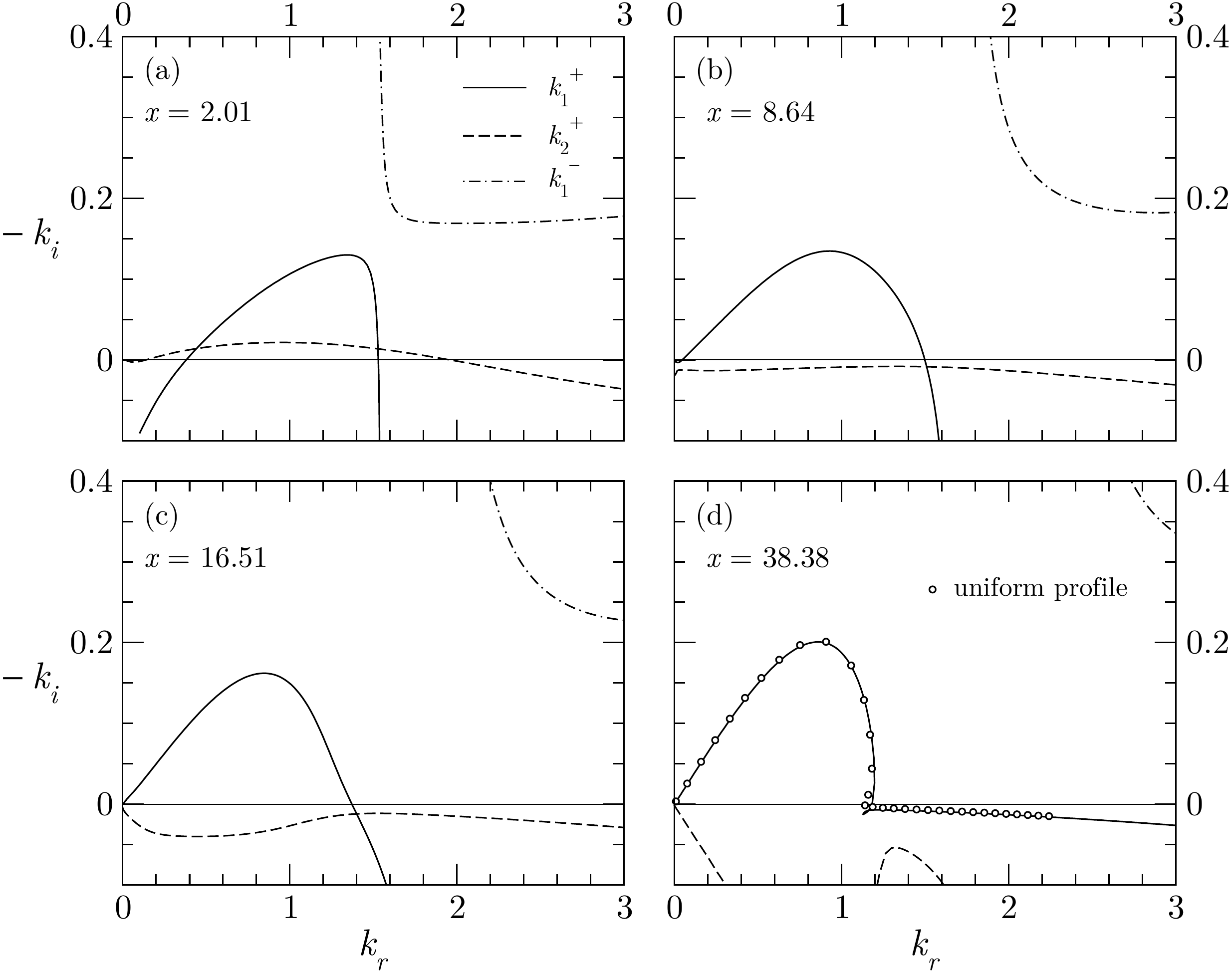}
    \caption{\label{fig:krki_G0_We310} Evolution of the spatial branches for $\Rey=200$, 
$\Web=3.1$, and different downstream positions along the jet: (a) $x=2.01$, (b) $x=8.64$, 
(c) $x=16.51$ and (d) $x=38.38$. The hollow circles in panel (d) correspond to the unstable 
branch of the Rayleigh-Chandrasekhar dispersion relation for a jet with uniform velocity 
profile.}
\end{center}
\end{figure}
The spatial instability modes evolve continuously along the jet. Thus, for instance,
figure~\ref{fig:krki_G0_We310} shows the spatial branches for $\Web=3.1$ and $\Rey=200$, 
performed at four different values of $x$, namely (a) $x=2.01$, (b) $x=8.64$, (c) $x=16.51$ 
and (d) $x=38.38$, with panel (a) being identical to figure~\ref{fig:modes}(e). Notice that 
the $k_1^+$ mode evolves downstream to eventually become the standard unstable branch of 
the Rayleigh-Chandrasekhar viscous dispersion relation in the developed region of the jet, 
i.e. for $X\gtrsim\mathcal{O}(10^{-1})$. In effect, the hollow circles in
figure~\ref{fig:krki_G0_We310}(d), computed with the viscous dispersion relation for a jet 
with uniform velocity profile, are seen to lie on top of the $k_1^+$ mode at the station 
$x=38.38$. Please also note that the wavenumber $k$ has been made dimensionless with the 
injector radius, $a$, instead of the local radius of the jet $aH(X)<a$, which explains the 
shift towards higher $k$ observed in figure~\ref{fig:krki_G0_We310}(d) in both the maximum 
and cut-off wavenumbers, with respect to their canonical counterparts. In contrast with the 
$k_1^+$ mode, the $k_2^+$ branch becomes stable sufficiently far downstream along the 
relaxation region, as expected from the fact that the base flow shear decreases as $x$ 
increases. Note that for the particular parameter values $\Web=3.1,\,\Rey=200$ the $k_2^+$ 
mode is already stable at $x=8.64$, as can be seen in figure~\ref{fig:krki_G0_We310}(b). In 
fact, irrespective of the values of $\Rey$ and $\Web$, this mode always stabilises at 
sufficiently large values of $x$ for which the base flow shear becomes small enough. Finally, 
it is important to point out that the instability is of convective type everywhere along the 
jet for $\Web=3.1$ and $\Rey=200$, since no pinching between downstream $(k^+)$ and upstream 
$(k^-)$ branches takes place at any value of $x$. Consequently, the signaling problem is well 
posed in this case, and the $k_1^+$ branch could then be used to describe the exponential 
downstream amplification of small disturbances leading to the breakup of the jet at some 
distance from the outlet.\\

\begin{figure}
\begin{center}
    \includegraphics[width=0.9\textwidth]{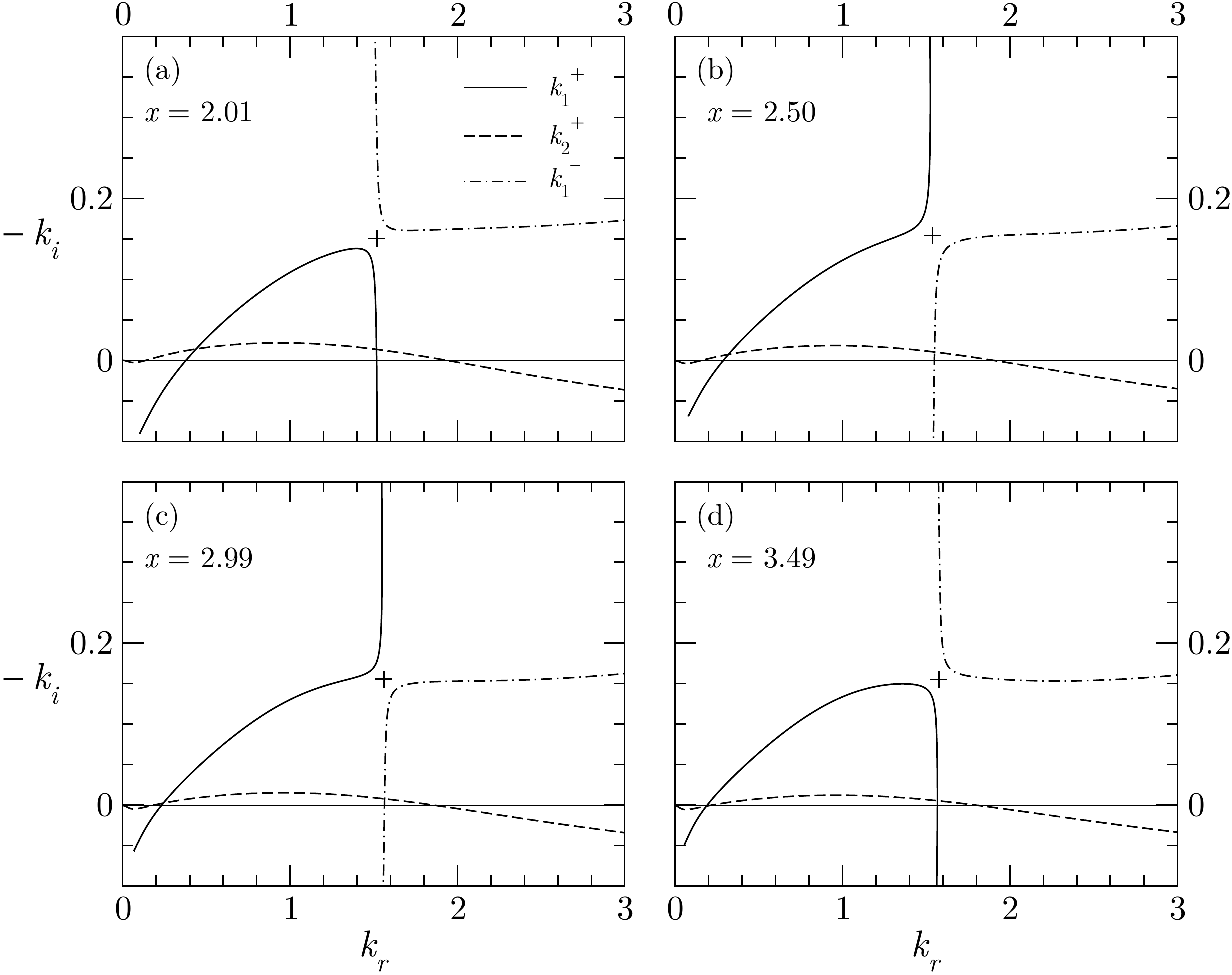}
    \caption{\label{fig:krki_G0_We300} Dowsntream evolution of the spatial branches for 
$\Rey=200$, $\Web=3$ and different downstream positions along the jet: (a) $x=2.01$, 
(b) $x=2.50$, (c) $x=2.99$ and (d) $x=3.49$. The cross marks the position of the 
double root, computed with the method of \cite{Deissler1987}.}
\end{center}
\end{figure}

Figure~\ref{fig:krki_G0_We300} displays the results of a spatial analysis performed for the 
same value of $\Rey=200$, but a slightly smaller value of $\Web=3$, for which a different 
scenario emerges. Thus, figure~\ref{fig:krki_G0_We300}(a) shows that, at $x=2$, the $k_2^+$ 
branch remains practically unchanged with respect to the case with $\Web=3.1$, the $k_1^+$ 
branch gives rise to a convective instability within a certain range of frequencies, and the 
$k_1^-$ branch has approached the $k_1^+$ branch. This behaviour persists along the jet down 
to the station $x_{ca}\simeq 2.11$, where the $k_1^+$ and $k_1^-$ solutions touch at a double 
root, $(k_0\simeq 1.525-0.151\text{i},\omega_0\simeq 0.682+0\text{i})$, of the dispersion 
relation, indicating the switch from a locally convective to a locally absolute instability 
in the jet at this particular position. Figures~\ref{fig:krki_G0_We300}(b) 
and~\ref{fig:krki_G0_We300}(c) show the structure of the $k$-plane at $x=2.5$ and $x=2.99$, 
downstream of the critical station where the mode interaction takes place. Note that $k_1^+$ 
and $k_1^-$ branches have exchanged identities as a consequence of their interaction in the 
neighborhood of the double root. More precisely, the spatial stability analysis becomes ill 
defined in this region since the different spatial branches can no longer be attributed to 
the downstream or upstream response of the flow. Instead, the local instability is absolute 
in this region due to the direct resonance between the $k_1^+$ and $k_1^-$ modes, giving 
rise to exponential amplification of perturbations both upstream and downstream from their 
source~\cite[for a detailed discussion, see][]{Huerre1990,Chomaz2005}. For the combination 
of parameters of figure~\ref{fig:krki_G0_We300}, the local absolute instability persists down 
to the station $x_{ac}\simeq 3.28$, where it becomes convective again, and so it remains for 
$x>x_{ac}$. An example is shown in figure~\ref{fig:krki_G0_We300}(d), calculated for $x=3.49$, 
where it can be seen that the $k_1^+$ and $k_1^-$ branches have interacted again at a double 
root, and recover the physical meaning they had in the region $x<x_{ca}$. A similar study of 
the $k$-plane was performed for the same value of the Reynolds number, $\Rey=200$, and 
different values of $\Web$, showing a similar sequence of events whenever 
$\Web\lesssim 3.07$. However, for $\Web\gtrsim 3.07$ the analysis revealed that the double 
root did not take place anywhere along the jet, indicating that the instability is convective 
everywhere in this case. Since, in the absence of feedback, the appearance of an unstable 
global mode requires the existence of a region of local absolute instability, this means 
that the jet with $\Rey=200$ should be globally stable for $\Web\gtrsim 3.07$.\\

\begin{figure}
\begin{center}
    \includegraphics[width=0.9\textwidth]{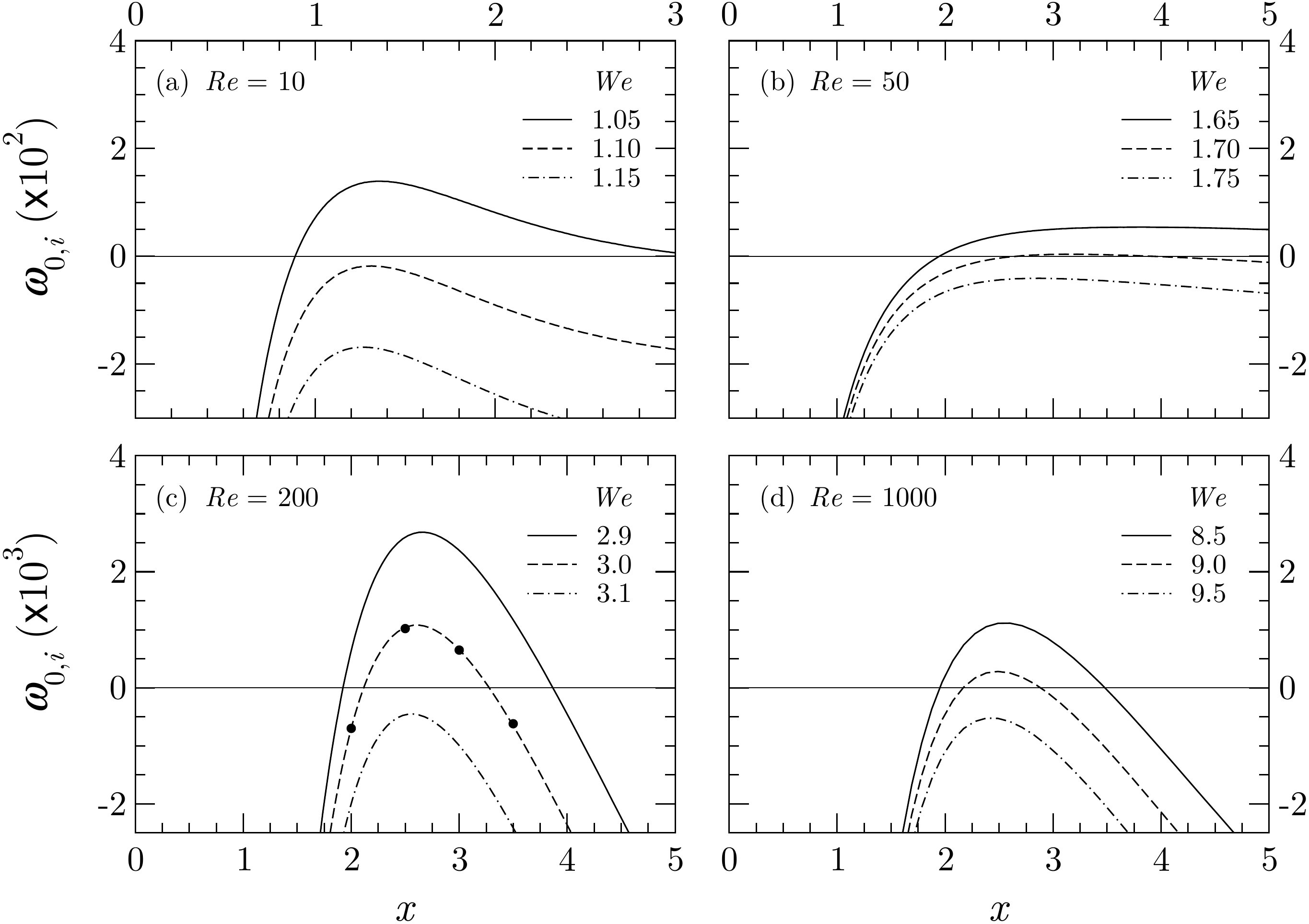}
    \caption{\label{fig:AGR_vs_x_G0} Downstream evolution of the absolute growth rate, 
$\omega_{0,i}(x)$, for $\Rey=(10,50,200,1000)$ in panels (a)-(d), respectively, and several 
values of $\Web$ indicated in the legends. The dots in panel (c) indicate the positions 
corresponding to figure~\ref{fig:krki_G0_We300}.}
\end{center}
\end{figure}
In order to extend the analysis to other values of $\Rey$, while avoiding the need for a 
complete characterization of the complex $k$-plane, we implemented the method developed 
by~\cite{Deissler1987} for the direct computation of double roots, $(k_0,\omega_0)$, of the 
dispersion relation. In particular, the crosses plotted in figure~\ref{fig:krki_G0_We300} 
were computed following this method. Once the double root was converged for fixed values of 
$(\Rey,\Web,x)$, it was followed in parameter space by means of a standard continuation 
method. Several examples of these computations can be seen in figure~\ref{fig:AGR_vs_x_G0}, 
which shows the downstream evolution of the \emph{absolute growth rate}, $\omega_{0,i}(x)$, 
for several values of the Reynolds and Weber numbers, as indicated in the panels. The dots 
in figure~\ref{fig:AGR_vs_x_G0}(c) correspond to the four downstream positions analysed in
figure~\ref{fig:krki_G0_We300}. Figure~\ref{fig:AGR_vs_x_G0} shows that the local instability
is convective near the outlet, $\omega_{0,i}(x=0)<0$, independently of the values of $\Rey$ 
and $\Web$. Therefore, the critical station $x_{ca}$ at which the instability type changes 
from convective to absolute, $\omega_{0,i}(x_{ca})=0$ with $d\omega_{0,i}/dx>0$, always 
takes place downstream of the outlet, $x_{ca}>0$. Moreover, figure~\ref{fig:AGR_vs_x_G0} 
reveals that, for a fixed value of $\Rey$, the absolute growth rate $\omega_{0,i}$ decreases 
as $\Web$ increases, as expected from physical grounds. For instance, 
figure~\ref{fig:AGR_vs_x_G0}(a) shows that, for $\Rey=10$ and $\Web=1.05$ (solid line), a 
region of absolute instability, $\omega_{0,i}>0$, exists within the jet. However, for a 
slightly larger value of $\Web=1.1$ (dashed line), the jet is locally convectively unstable 
throughout the domain, $\omega_{0,i}(x)<0\,\forall x$. Consequently, if the critical Weber 
number, $\Web_c$, is defined according to the condition $max[\omega_{0,i}(x;\Web_c,\Rey)]=0$, 
it is such that $1.05\leq \Web_c \leq 1.1$ for $\Rey=10$. A similar trend can be seen to 
hold for the other Reynolds numbers shown in figure~\ref{fig:AGR_vs_x_G0}, with the 
corresponding values of $\Web_c$ increasing with $\Rey$. Thus, from 
figures~\ref{fig:AGR_vs_x_G0}(b)-\ref{fig:AGR_vs_x_G0}(d) it is deduced that 
$1.7\leq \Web_c(\Rey=50) \leq 1.75$, $3\leq \Web_c(\Rey=200) \leq 3.1$, and 
$9\leq \Web_c(\Rey=1000) \leq 9.5$, respectively. To obtain more precise values of the 
function $\Web_c(\Rey)$, the curves were computed for small increments of $\Rey$ in the 
range $10\leq \Rey \leq 1000$, and the corresponding values of $\Web_c$ were then obtained
by a standard root finding technique together with a continuation method.\\

\begin{figure}
\begin{center}
    \includegraphics[width=0.9\textwidth]{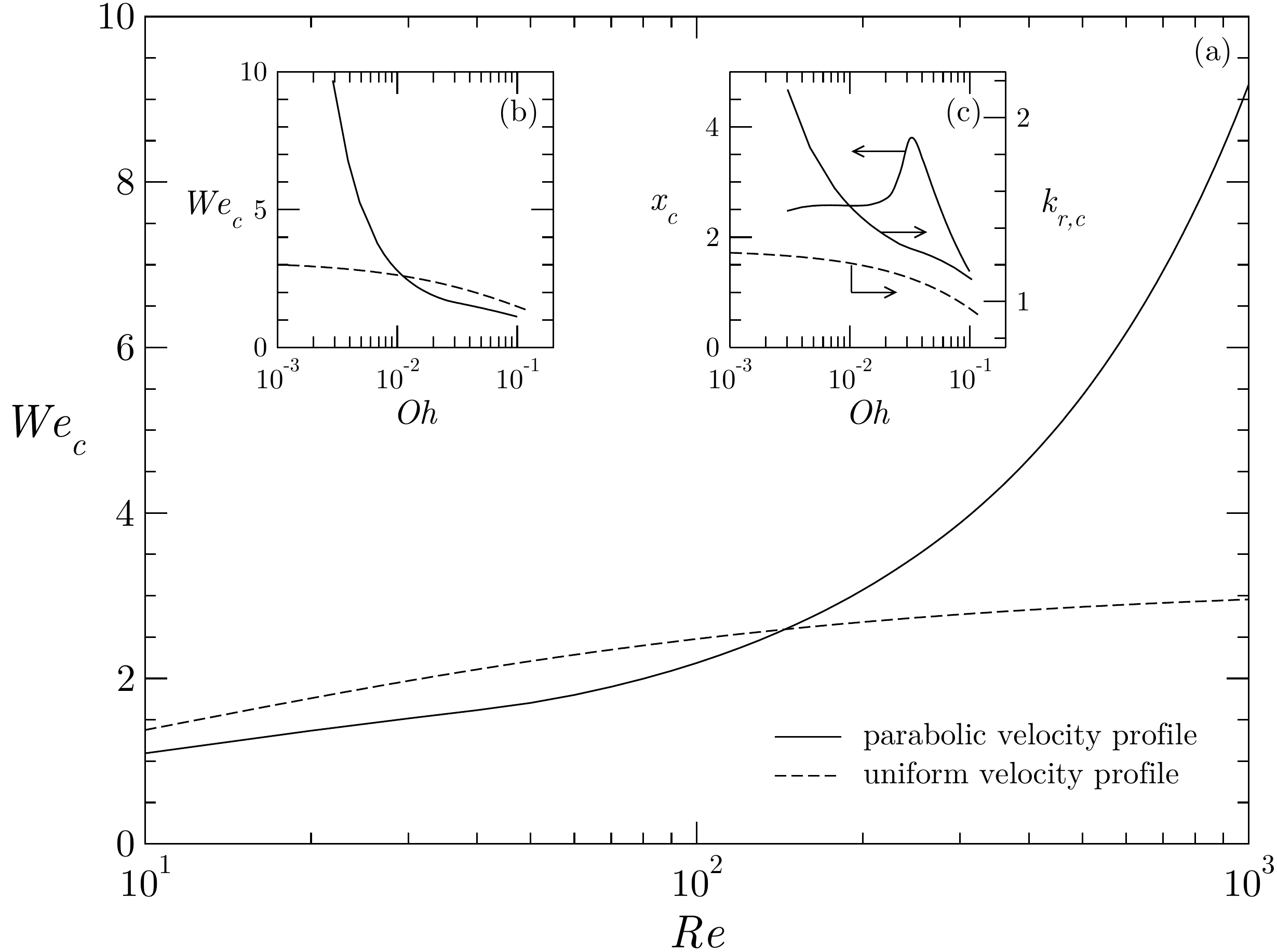}
    \caption{\label{fig:Wec_vs_Re_G0} (a) Critical Weber number, $\Web_c$, as a function 
of $\Rey$ for the onset of absolute instability without gravity, $G=0$. (b) The function 
$\Web_c(\Oh)$ and (c) the functions $x_c(\Oh)$ (left axis) and $k_{r,c}(\Oh)$ (right axis).
The solid line shows the results obtained for parabolic velocity profile at the outlet, 
while the dashed line corresponds to uniform velocity profiles~\citep[][]{LeibyGoldsteinAC}.}
\end{center}
\end{figure}

Figure~\ref{fig:Wec_vs_Re_G0} summarises the results obtained in the present section for
the case without gravity, $G=0$. Thus, the dependence of the critical Weber number on the 
Reynolds number is shown in figure~\ref{fig:Wec_vs_Re_G0}(a), both for a jet with uniform
velocity profile, obtained by~\cite{LeibyGoldsteinAC} (dashed line), and for a jet with 
fully-developed profile at the outlet considered in the present work (solid line). Both 
results are seen to coincide at the point $(\Rey,\Web)\simeq (145,2.6)$. On the one hand, 
for $\Rey<145$, the jet with parabolic outlet profile is slightly more stable than the 
uniform profile jet, in the sense that the former has smaller associated values of 
$\Web_c$ than the latter. On the other hand, for $\Rey>145$, while the uniform profile 
jet tends monotonically to the inviscid asymptote $\Web_c\to 3.1$ as $\Rey\to\infty$ 
\cite[][]{LeibyGoldstein}, the slope of the curve $\Web_c(\Rey)$ is seen to increase 
with $\Rey$. Note that, as mentioned in \S\ref{sec:intro}, an upper limit of $\Rey=1000$ 
has been set in the computations in order to ensure laminar flow in the jet. Nevertheless,
notice that both results differ substantially in the range $200\lesssim\Rey\lesssim 1000$,
with the parabolic jet reaching values as high as $\Web\simeq 9.25$ at $\Rey=1000$. 
Unfortunately, although several authors have reported experimental results of the 
jetting/dripping transition under microgravity or reduced gravity conditions using 
nozzles \cite[see for instance][]{Vihinen1997,Odonnell2001,Osborne2006}, to the best of 
the author's knowledge no data is available in the literature for low Ohnesorge number
capillary jets discharging from long needles in this case. Consequently, new experiments
would be needed in order to check the results of figure~\ref{fig:Wec_vs_Re_G0}. It is
important to emphasise the strong differences between our results and those obtained 
by~\cite{LeibyGoldstein} in their inviscid spatiotemporal analysis of the parametric 
family of profiles $(1-br^2)/(1-b/2),\;0\leq b\leq 1$, aimed at modeling the effects 
of viscous relaxation. In particular, figure~9 of~\cite{LeibyGoldstein}, which shows 
$\Web_c$ as a function of the parameter $b$, gives a range of values for the critical 
Weber number going from $\Web_c\simeq 0.45$ for $b=1$, corresponding to the parabolic 
Poiseuille profile, to $\Web_c\simeq 3.1$ for $b=0$, i.e. the inviscid asymptote for 
the uniform profile, having a local maximum $\Web_c\simeq 3.24$ at $b\simeq 0.14$. 
Although \cite{LeibyGoldstein} argue that the effect of the downstream evolution of the 
jet would correspond to a parametric movement from $b=1$ to $b=0$, it is clear that the 
value of $\Web_c<3.25$ independently of the relationship between the parameter $b$ and 
the downstream position $x$, leading to results completely different from those shown
in figure~\ref{fig:Wec_vs_Re_G0}(a). Since the values of $\Rey$ and $\Web$ are linked 
through the Ohnesorge number by $\Oh=\Web^{1/2}/\Rey$, the critical conditions can 
alternatively be described, without loss of generality, by the function $\Web_c(\Oh)$ 
represented in figure~\ref{fig:Wec_vs_Re_G0}(b), whose usefulness lies in the fact that 
$\Oh$ is independent of the jet velocity. In addition, we have also plotted in
figure~\ref{fig:Wec_vs_Re_G0}(c) the dependence of the real part of the absolute 
wavenumber evaluated at the critical conditions ($k_{r,c}$, right axis), as well as the 
station $x_c$ where marginal local absolute instability takes place (left axis).
Figure~\ref{fig:Wec_vs_Re_G0}(c) reveals, in particular, that the absolute mode 
associated to the uniform profile jet (dashed line) has a longer wavelength compared to 
that of a jet with parabolic outlet profile (solid line). Moreover, in the latter case 
the marginal station is seen to present a non-monotonic behaviour, varying in the range 
$1.5\lesssim x_c \lesssim 4$, with a low-$\Oh$ plateau at $x_c\simeq 2.5$.\\

\begin{figure}
\begin{center}
    \includegraphics[width=0.9\textwidth]{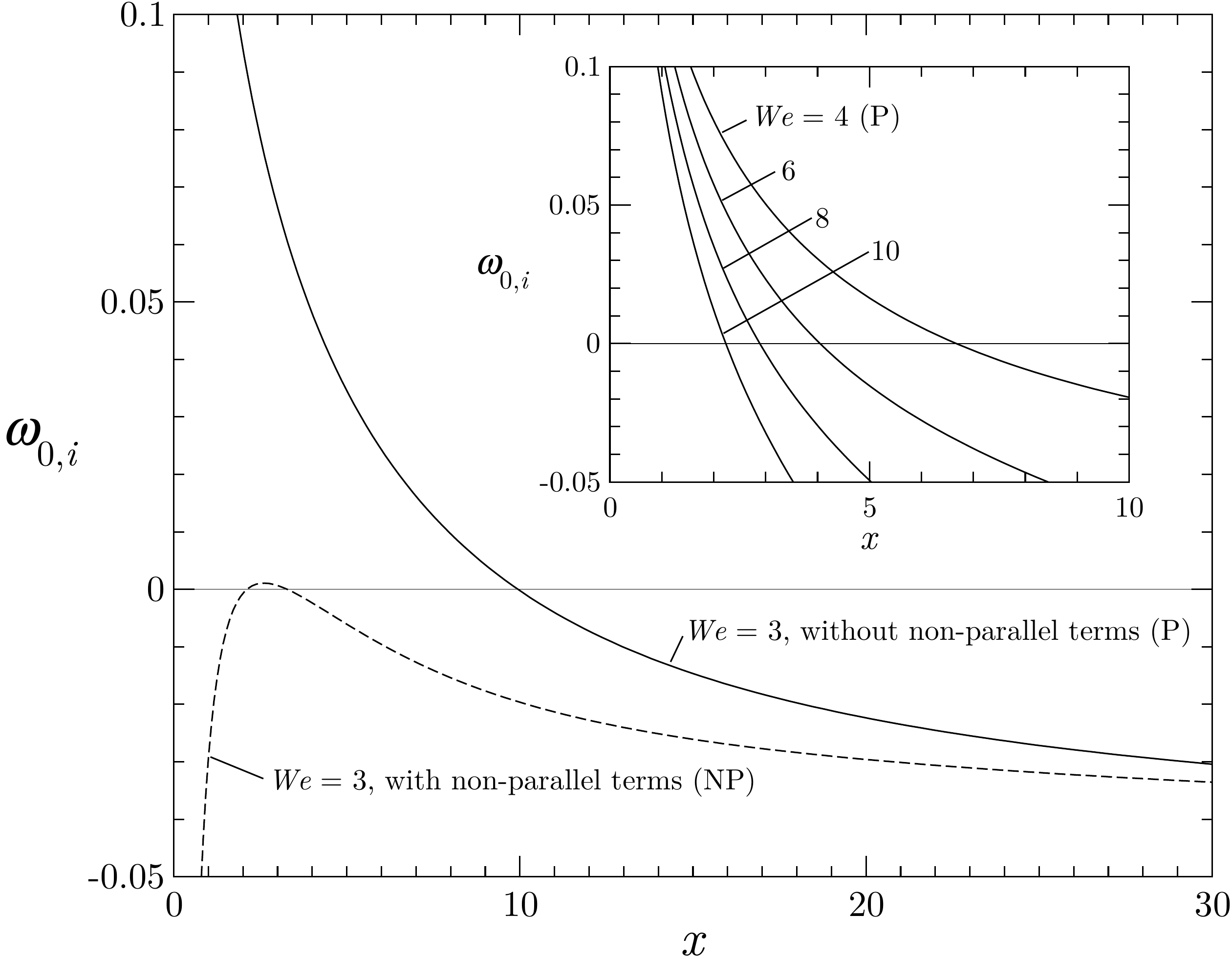}
    \caption{\label{fig:agrp} Downstream evolution of $\omega_{0,i}$ for $\Rey=200$ and 
$\Web=3$, computed with (case NP, dashed line) and without (case P, solid line) non-parallel 
terms in the stability equations. The inset shows the function $\omega_{0,i}(x)$, calculated 
for the case P, for $\Rey=200$ and values of $\Web=(4,6,8,10)$, which predicts a region of 
absolute instability close to the nozzle regardless of the value of $\Web$.}
\end{center}
\end{figure}
The picture described above changes completely when non-parallel terms are disregarded 
in the stability equations, i.e. when the right hand sides of 
equations~(\ref{eq:pse2})-(\ref{eq:bcpse3}) are neglected. This approximation will hereafter
be referred to as \emph{case P}, in contrast with the computations performed retaining the 
non-parallel terms, which will be called \emph{case NP}. Figure~\ref{fig:agrp} shows the 
function $\omega_{0,i}(x)$ for $\Rey=200$ and $\Web=3$, calculated both for the P and NP 
cases (solid and dashed lines, respectively). Figure~\ref{fig:agrp} reveals that both 
approaches give markedly different results for values of $x\lesssim 20$, i.e. values of
$X\lesssim\mathcal{O}(0.1)$, for which non-parallel terms give order unity contributions 
to the stability properties at moderately large values of $\Rey$, as was previously noted. 
In particular, the values of $\omega_{0,i}$ in case P are larger than in case NP all along 
the jet. Since far downstream the jet velocity profile tends to be uniform, and its radius 
tends to a constant value, both results tend to a common limit as $x\to\infty$. However, 
the crucial difference lies in their behaviour near the outlet. As can be observed in 
figure~\ref{fig:agrp}, in case P the value of $\omega_{0,i}$ increases monotonically as 
$x\to 0$. In effect, the value of the absolute growth rate at a given jet station results 
from a competition between the degree of instability and the convective velocity of the 
growing disturbances. Since the eigenfunctions of the $k_1^+$ and $k_1^-$ modes peak at the 
interface [see figure~\ref{fig:modes}], it is expected that the decrease in surface velocity 
which takes place as $x\to 0$ leads to an increase in the values of $\omega_{0,i}$. 
Although the same reasoning applies to the NP model, in this case non-parallel terms 
introduce a \emph{stabilising} influence which compensates the destabilisation associated 
with the reduced free-surface velocity. For instance, note that the non-parallel convective 
term in the $x$-momentum equation~(\ref{eq:pse2}), $U_X\,\hat{u}$, enhances the dowsntream 
transport of the perturbations, thus leading to the reduced values of $\omega_{0,i}$ observed 
in figure~\ref{fig:agrp} for case NP, when compared to case P. Moreover, this stabilising 
effect is enhanced as $x$ decreases, as a consequence of the fact that the acceleration of 
the base flow at the interface, $U_X(X,r=H)$, increases very fast as $X\to 0$. In particular, 
as mentioned before, $\omega_{0,i}<0$ at the outlet, and the locally absolutely unstable 
region always takes place inside the jet in case NP. In addition, the inset of 
figure~\ref{fig:agrp} shows that, although in case P the size of the absolutely unstable 
region decreases as $\Web$ increases, it does always exist sufficiently close to the exit 
and, in particular, $\omega_{0,i}>0$ at $x=0$ irrespective of the value of $\Web$. It is 
worth mentioning that a behaviour similar to the case P was found in the case of planar 
liquid sheets undergoing viscous relaxation by~\cite{Soderberg03}, where non-parallel terms 
were discarded in the stability equations, and a region of absolute instability was shown 
to be present close to the nozzle. As happens here, although the size of the absolutely 
unstable region decreases as $\Web$ increases, it does not lead to a clear identification 
of a globally stable jet. Moreover, this behaviour poses a fundamental problem when considering 
the spatial instability of the jet, not specifically addressed here, which would always be 
ill-defined due to the presence of the absolutely unstable region near the outlet. In the 
remainder of the paper, the analysis will be performed retaining the non-parallel terms in 
the linear stability equations.

\subsection{Results for finite $G$: jetting/dripping transition in water jets}
\label{subsec:water}

Let us now present several results obtained taking gravity into account. As pointed out 
by~\cite{ledizes97}, the parameters of the problem, $\Web,\Rey$ and $\text{G}$, are not 
independent from each other given the values of the liquid properties $\rho,\nu$ and 
$\sigma$, and the gravitational acceleration $g$. In looking for an alternative set of
parameters that enables a more straightforward comparison with experimental results, it
is convenient to express the dependence on the length scale through the Bond number,
$\Bo=\rho ga^2/\sigma$. The values of the Stokes and Reynolds numbers, needed in the 
computation of the basic flow and its stability, are then uniquely specified in terms 
of the Weber number as $\text{G}=(\Bo^5/\Mo)^{1/4}\Web^{-1/2}$ and 
$\Rey=(\Bo/\Mo)^{1/4}\Web^{1/2}$, where $\Mo=\nu^4 \rho^3 g/\sigma^3$ is the Morton 
number of the liquid, independent of $a$ and $Q$. As a result, the critical Weber number 
for the appearance of absolute instability in the jet can be expressed alternatively as 
a function of the form $\Web_c\left(\Bo,\Mo\right)$, which for a given liquid simplifies 
to $\Web_c\left(\Bo\right)$. The results presented in this section have been computed for 
water at room temperature, for which $\Mo\simeq 2.52\times 10^{-11}$, corresponding to the 
experiments performed by \cite{Clanet1999}. Notice that, in the particular case of water, 
the Stokes number is $\text{G}\simeq 446\,\Bo^{5/4}\Web^{-1/2}$ and, since it will be seen 
that $\Web_c\lesssim\mathcal{O}(1)$, the effect of gravity cannot be neglected except for 
sufficiently small injector diameters, $\Bo\lesssim \mathcal{O}(10^{-2})$ (see 
figure~\ref{fig:Wec_vs_Bo}).\\

\begin{figure}
\begin{center}
    \includegraphics[width=0.9\textwidth]{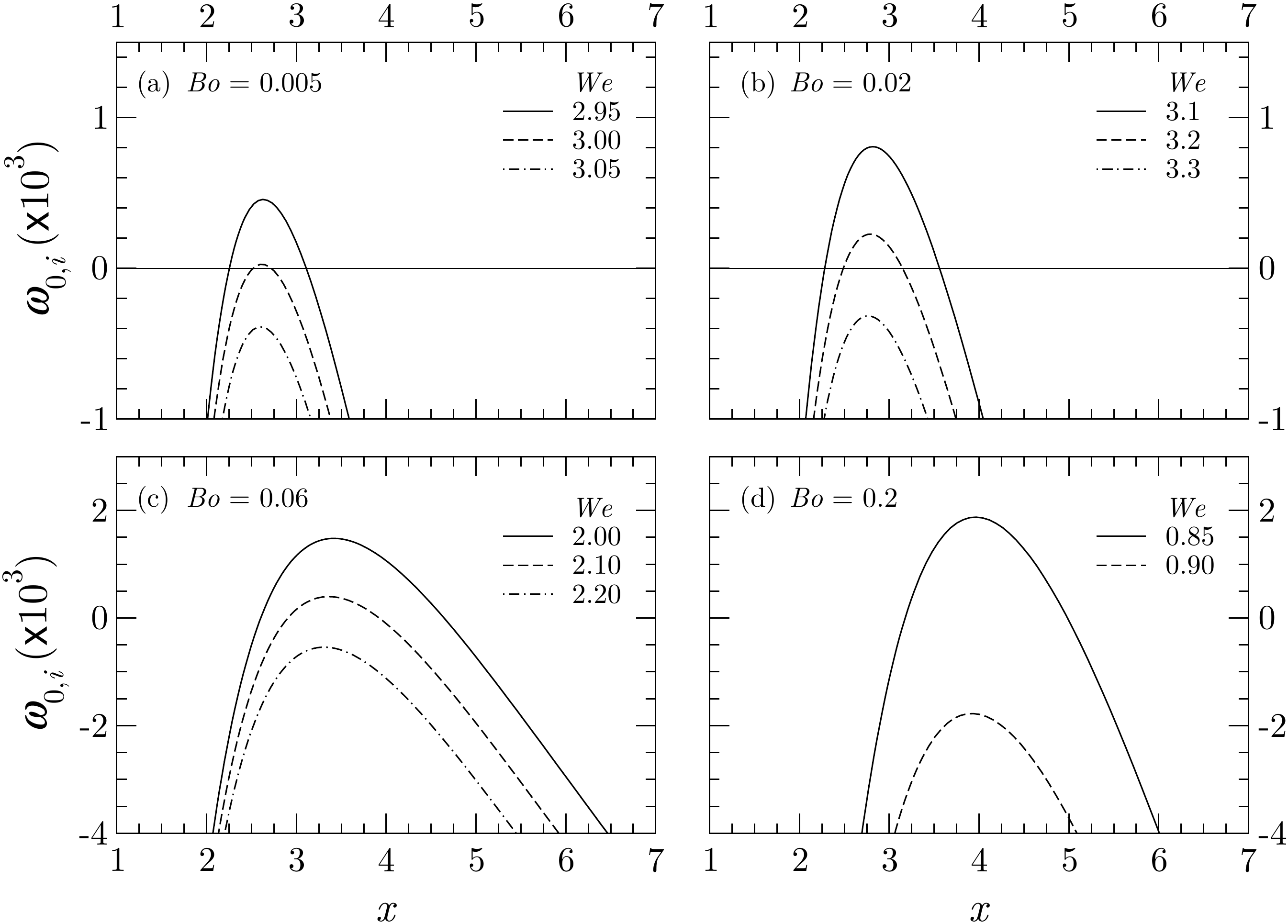}
    \caption{\label{fig:AGR_vs_x} Downstream evolution of the absolute growth rate, 
$\omega_{0,i}(x)$, for $\Mo=2.52\times 10^{-11}$, corresponding to water at room 
temperature, and (a) $\Bo=0.005$, (b) $\Bo=0.02$, (c) $\Bo=0.06$ and (d) $\Bo=0.2$,
and different values of $\Web$ indicated in the legends.}
\end{center}
\end{figure}
The spatiotemporal analysis presented in this section follows the same lines as that 
of~\S\ref{subsec:G0}. It is important to emphasise that, although the local stability 
properties obtained for the jet with $G>0$ present quantitative differences from those 
presented in \S\ref{subsec:G0} for the case $G=0$, the phenomenology is qualitatively
similar. In particular, the spatiotemporal stability properties are governed by the 
interaction of the same spatial instability modes described in \S\ref{subsec:G0}, namely 
$k_1^+$ and $k_1^-$. Consequently, in this section we will directly present computations
of the double root $(k_0,\omega_0)$. Specifically, for given values of $\Bo$ and $\Web$, 
$(k_0,\omega_0)$ was computed as a function of $x$ by means of a continuation method, with 
the values of $G$ and $\Rey$ obtained as described in the previous paragraph, with the aim 
at obtaining the function $\Web_c(\Bo)$ for water jets. The results obtained are 
summarised in figures~\ref{fig:AGR_vs_x} and~\ref{fig:Wec_vs_Bo}.\\

The downstream evolution of the absolute growth rate, $\omega_{0,i}\,(x)$, is represented 
in figures~\ref{fig:AGR_vs_x}(a)-\ref{fig:AGR_vs_x}(d) for the values of 
$\Bo=(0.005,0.02,0.06,0.2)$, respectively, and several values of $\Web$ as indicated in 
the legends. Note that the shapes of the curves are similar to those shown in
figure~\ref{fig:AGR_vs_x_G0}: the maximum value $\omega_{0,i}^m$ takes place inside the 
jet at a station $x_m$ which varies from $x_m\simeq 2.5$ for $\Bo=0.005$, to $x_m\simeq 4$ 
for $\Bo=0.2$. Moreover, for each value of $\Bo$, $\omega_{0,i}^m$ decreases as $\Web$ 
increases. Thus, for instance, figure~\ref{fig:AGR_vs_x}(a) shows that, for $\Bo=0.005$ 
and $\Web=2.95$ (solid line), $\omega_{0,i>0}$ in the range $2.25\lesssim x \lesssim 3.12$. 
However, for $\Web=3$ (dashed line), the flow is only marginally absolutely unstable near 
$x\simeq 2.3$, and it is locally convectively unstable throughout for slightly larger value 
of $\Web=3.05$ (dash-dotted line). Consequently, the critical Weber number accomplishes 
$3<\Web_c<3.05$ for $\Bo=0.005$. A similar trend is observed in 
figures~\ref{fig:AGR_vs_x}(b)-\ref{fig:AGR_vs_x}(d), with the value of $\Web_c$ showing 
a non-monotonic behaviour for increasing values of $\Bo$. In effect, note that
$3.2<\Web_c(\Bo=0.02)<3.3$, $2.1<\Web_c(\Bo=0.06)<2.2$ and $0.85<\Web_c(\Bo=0.2)<0.9$.\\

\begin{figure}
\begin{center}
    \includegraphics[width=1.00\textwidth]{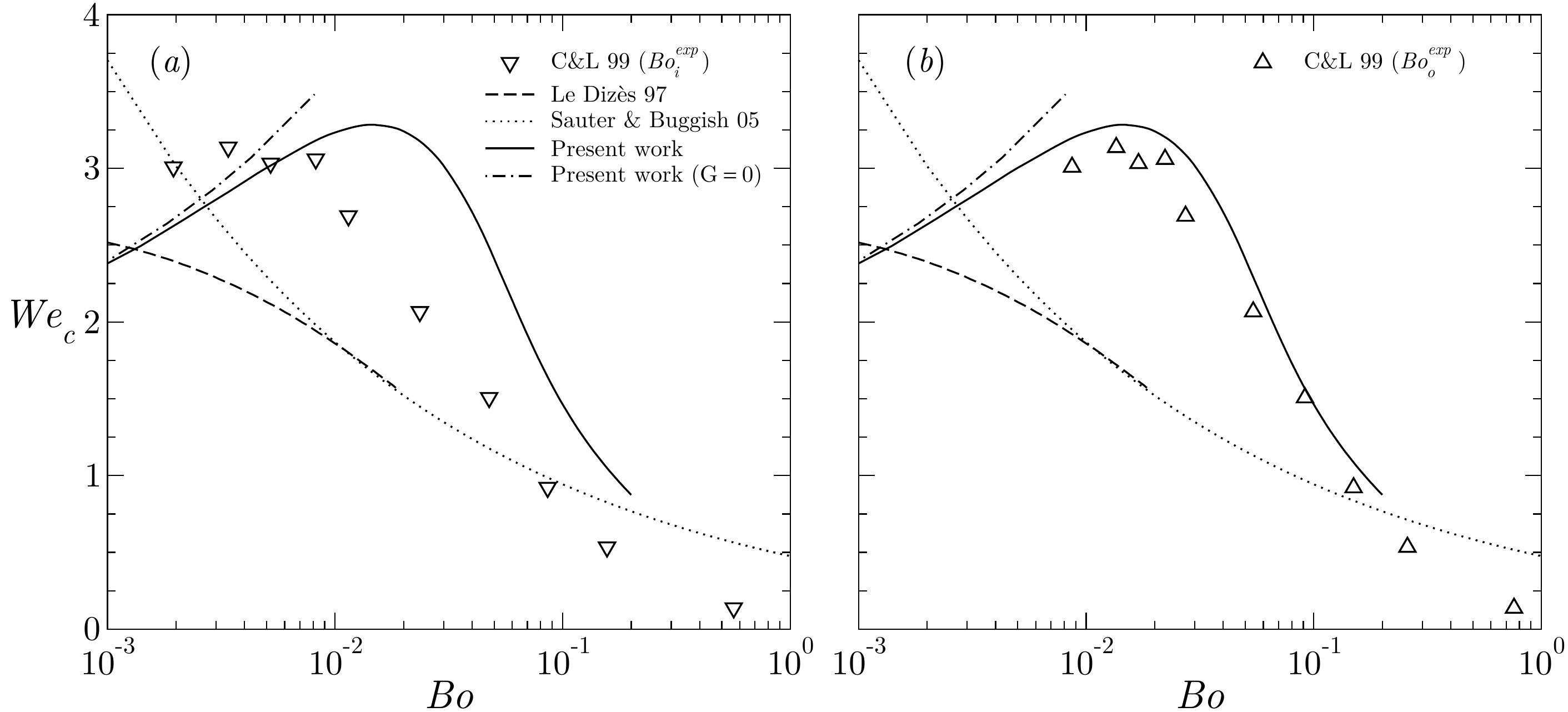}
    \caption{\label{fig:Wec_vs_Bo} The function $\Web_c(\Bo)$ for transition from 
jetting to dripping in water obtained from the experiments of~\cite{Clanet1999}, 
using (a) the inner needle radius ($\triangledown$), and (b) the outer needle radius 
($\vartriangle$), to compute the value of $\Bo$. Also shown are the results of the 
present work with and without gravity (solid and dash-dotted lines, respectively), 
as well as those by~\cite{ledizes97} (dashed line) and~\cite{SauteryBuggisch} 
(dotted line).}
\end{center}
\end{figure}

Following a procedure similar to that described in \S\ref{subsec:G0} for the case without 
gravity, a more detailed calculation was performed to obtain an accurate representation of 
the function $\Web_c(\Bo)$ in the range $0.001\leq \Bo \leq 0.2$. The upper limit in $\Bo$ 
was set in the computations due to the fact that the slenderness requirement 
$1\ll \Fr=\Web/\Bo$ is manifestly violated for $\Bo\gtrsim 0.2$. On the other hand, since
$\Rey=(\Bo/\Mo)^{1/4}\Web^{1/2}$, $\Web_c^{1/2}\gtrsim\mathcal{O}(1)$ and $\Bo\gg \Mo$,
the restriction $\Rey\gg 1$ is always accomplished in this case. Figure~\ref{fig:Wec_vs_Bo} 
shows the function $\Web_c(\Bo)$ obtained in the present work for the cases wihout gravity 
(dash-dotted line), and with gravity (solid line). Please note that the dash-dotted line 
in figure~\ref{fig:Wec_vs_Bo} corresponds to the function $\Web_c(\Rey)$ obtained 
in~\S\ref{subsec:G0}, with the Reynolds number given by $\Rey=(\Bo/\Mo)^{1/4}\Web^{1/2}$,
as explained above. The results were computed for a value of $\Mo=2.52\times 10^{-11}$, 
corresponding to water at room temperature. Also shown are the experimental results 
of~\cite{Clanet1999} for the transition from jetting to dripping in water jets discharging 
from needles long enough for the velocity profile to be fully developed at the outlet 
($\triangledown,\vartriangle$). As mentioned in~\S 2, these experiments were conducted with 
hypodermic needles having a finite wall thickness, and the contact line was always pinned at 
the \emph{outer} edge of the exit section. Therefore, an unambiguous comparison between theory 
and experiments is precluded by the fact that the flow configuration is not exactly the same 
as that considered in the present work, which only considers the case of a vanishingly thin 
wall thickness. In particular, to compute the values of the Weber and Bond numbers associated 
to the experimental data points obtained by~\cite{Clanet1999}, one may arbitrarily choose 
either the inner or the outer radius as characteristic length scale. Note 
that~\cite{SauteryBuggisch}, who also disregarded finite wall thickness effects in their 
analysis, faced the same problem when comparing their results with those by~\cite{Clanet1999}. 
In particular, \cite{SauteryBuggisch} chose the inner diameter of the nozzle to obtain the 
experimental values of both $\Web$ and $\Bo$. Here, we shall also choose the inner radius to 
compute the experimental value of the Weber number, 
$\Web^{\text{exp}}=\rho\,Q_{\text{exp}}^{2}/(\pi^2\sigma\,a_i^3)$, where $Q_{\text{exp}}$ is 
the experimental volume flow rate and $a_i$ is the inner needle radius. Regarding the Bond 
number, we decided to include two different data series, where $\Bo$ is based either on the 
inner or on the outer radius. Thus, figures~\ref{fig:Wec_vs_Bo}(a) and~\ref{fig:Wec_vs_Bo}(b) 
respectively make use of the expressions $\Bo_{i}^{\text{exp}}=\rho g a_i^2/\sigma$ and 
$\Bo_{o}^{\text{exp}}=\rho g a_o^2/\sigma$ for the Bond number, where $a_o$ represents the 
outer radius. In addition, figure~\ref{fig:Wec_vs_Bo} also includes the results obtained 
by~\cite{ledizes97} and~\cite{SauteryBuggisch}, hereafter referred to as LD97 and S\&B05, 
for liquid jets with uniform velocity profiles (dashed and dotted lines, respectively). 
The results of LD97 and S\&B05 coincide in the range $0.009\lesssim\Bo\lesssim 0.02$, but 
separate from each other for $\Bo\lesssim 0.009$, the values of $\Web_c$ given by S\&B05 
being larger than those of LD97 in that range. Nevertheless, both curves are seen to 
underestimate the experimental values of $\Web_c$ up to $\Bo\simeq 0.1$, where the result 
of S\&B05 crosses the experimental data, overestimating them for $\Bo\gtrsim 0.1$. In contrast, 
the results obtained in~\S\ref{subsec:water} of the present work (solid line) are seen to 
provide better overall agreement with the experimental trend in a wide range of Bond numbers, 
especially in the case of figure~\ref{fig:Wec_vs_Bo}(b), where the outer radius is used to 
compute the Bond number. The result obtained in~\S\ref{subsec:G0} for the buoyancy-free jet, 
$G=0$, (dash-dotted line) agrees with the full computation only for values of 
$\Bo\lesssim 0.005$. Thus, in addition to viscous relaxation, the effect of gravity must also 
be taken into account in the stability analysis to properly describe the transition from 
jetting to dripping observed in experiments.

\section{Conclusions}\label{sec:conclusions}
A spatiotemporal linear stability analysis has been performed with the aim of predicting the 
appearance of local absolute instability in capillary jets undergoing velocity profile 
relaxation at large Reynolds and Froude numbers. Although the reference case of a fully 
developed flow at the exit was analysed in detail, the approach presented in this work could 
also be used to study other exit velocity profiles, e.g. those corresponding to nozzles or short 
injection tubes. Nevertheless, it could well be that several conclusions drawn from the present 
work apply also to other configurations, independently of the exact shape of the initial 
velocity profile. Thus, for instance, we have highlighted the importance of retaining 
non-parallel terms in the local stability equations, by comparing computations performed with 
and without these terms. It is important to emphasise that non-parallel terms enter the local 
stability problem at the same order as viscous effects in the perturbations, namely $\Rey^{-1}$. 
Consequently, they can be neglected only when base flow variations and transverse velocities 
become small, as happens sufficiently far downstream from the outlet, where the computations 
performed with and without non-parallel terms coincide. However, it is precisely the region 
near the outlet where absolute growth rates become larger due to the small values of the free 
surface velocity. In fact, when calculated without non-parallel terms, the slope of the local 
absolute growth rate satisfies $d\omega_{0,i}(x)/dx<0$ as $x\to 0$. This behaviour is similar 
to that found by~\cite{Soderberg03} in his analysis of the relaxing two-dimensional liquid 
sheet for values of the Weber number smaller than one. In contrast, we have shown here that, 
in the axisymmetric case, a region of local absolute instability exists near the exit 
independently of the value of $\Web$. However, when non-parallel terms are considered, it has 
been found that $\omega_{0,i}<0,\,d\omega_{0,i}(x)/dx>0$ as $x\to 0$ independently of $\Web$ 
and, therefore, the local instability is always convective near the outlet. Thus, the strongly 
stabilising effect of non-parallel terms is able to compensate for the reduced convection velocity 
of disturbances. This fact is clearly seen, for instance, in the positive convective acceleration 
of the basic flow near the interface, $U_X>0$, which promotes an enhanced downstream transport 
of the axial velocity perturbations, $U_X\,\hat{u}$, when compared to the fully parallel 
approach and which, in turn, leads to smaller values of $\omega_{0,i}$. Notice that the last 
conclusion would hold irrespective of the exact shape of the velocity profile, since it relies 
only on the fact that the free surface accelerates near the outlet due to the substitution of 
the no-slip condition at the injector wall to a free surface boundary condition.

The formulation used in the present work leads to a natural prediction for the critical 
Weber number, $\Web_c$, below which a global mode is destabilised and subsequently destroys 
the liquid column leading to a dripping regime. In effect, $\omega_{0,i}(x)$ presents a local 
maximum inside the jet and, following~\cite{Pier98}, $\Web_c$ is defined according to the 
marginal condition that this maximum be zero. For values of $\Web>\Web_c$, the flow is 
convectively unstable throughout the jet, while, for $\Web<\Web_c$, a localised region of 
absolute instability takes place in its near field. The function $\Web_c(\Rey)$ has been 
computed for the case without gravity, showing marked differences with respect to the case 
of a jet with uniform velocity profile considered by~\cite{LeibyGoldsteinAC}. Unfortunately, 
the available experiments under microgravity conditions were performed either with 
nozzles~\citep[][]{Odonnell2001,Osborne2006} or at very small Reynolds 
numbers~\citep[][]{Vihinen1997}, thus precluding a comparison with our results. Given the 
liquid and the value of $g$, the functional dependence can be alternatively written 
$\Web_c(\Mo,\Bo)$, where $\Mo$ is the Morton number, which depends neither on the injector 
radius nor on the jet velocity, and $\Bo$ is the Bond number. The function $\Web_c(\Bo)$, 
computed here for the particular case of water jets under Earth gravity, provides better 
agreement with the experiments of~\cite{Clanet1999} than previous studies using uniform 
velocity profiles~\cite[][]{ledizes97,SauteryBuggisch}, thus giving support to the approach 
used in the present work. It is important to emphasise that, implicit to this approach, is 
the assumption that the appearance of a pocket of absolute instability leads to the 
destabilisation of a global mode, independently of the extent of the absolutely unstable 
domain. Positive evidence for this fact has been given in different contexts, going from 
model one-dimensional problems~\cite[][]{Pier98} to actual Navier-Stokes 
flows~\cite[see for instance][]{Pier2001,Sevilla2004,Lesshafft2006}. Finally, let us point 
out that a more rigorous approach to analyse the flow under consideration would require 
either using BiGlobal stability analysis~\cite[][]{Theofilis2011}, or performing direct 
numerical simulations. These tasks are left for future work.


\begin{acknowledgments}
Financial support by the Spanish Ministry of Education, Comunidad de Madrid and Universidad 
Carlos III de Madrid under projects DPI2008-06624-C03-02 and CCG10-UC3M/DPI-4777 is acknowledged.
This paper is devoted to the memory of Professor Antonio Barrero, whose integrity will perdure.
\end{acknowledgments}




\begin{thebibliography}{50}
\expandafter\ifx\csname natexlab\endcsname\relax\def\natexlab#1{#1}\fi

\bibitem[Ambravaneswaran {\em et~al.\/}(2004)Ambravaneswaran, Subramani,
  Phillips \& Basaran]{Ambravaneswaran2004}
{\sc Ambravaneswaran, B., Subramani, H.~J., Phillips, S.~D. \& Basaran, O.~A.}
  2004 Dripping-jetting transitions in a dripping faucet. {\em Phys. Rev.
  Lett.\/} {\bf 93}, 034501.

\bibitem[Barrero \& Loscertales(2007)]{BarreroAR}
{\sc Barrero, A. \& Loscertales, I.~G.} 2007 Micro- and nanoparticles via
  capillary flows. {\em Annu. Rev. Fluid Mech.\/} {\bf 39}, 89--106.

\bibitem[Basaran(2002)]{Basaran2002}
{\sc Basaran, O.~A.} 2002 Small-scale free surface flows with breakup: drop
  formation and emerging applications. {\em AIChE J.\/} {\bf 48}, 1842--1848.

\bibitem[Bogy(1979)]{Bogy}
{\sc Bogy, D.~B.} 1979 Drop formation in a circular liquid jet. {\em Annu. Rev.
  Fluid Mech.\/} {\bf 11}, 207--228.

\bibitem[Chomaz(2005)]{Chomaz2005}
{\sc Chomaz, J.-M.} 2005 Global instabilities in spatially developing flows:
  Non-normality and nonlinearity. {\em Ann. Rev. Fluid Mech.\/} {\bf 37},
  357--392.

\bibitem[Chomaz {\em et~al.\/}(1988)Chomaz, Huerre \& Redekopp]{Chomaz1988}
{\sc Chomaz, J.-M., Huerre, P. \& Redekopp, L.~G.} 1988 Bifurcation to local
  and global modes in spatially developing flows. {\em Phys. Rev. Lett.\/} {\bf
  60}, 25--28.

\bibitem[Clanet \& Lasheras(1999)]{Clanet1999}
{\sc Clanet, C. \& Lasheras, J.~C.} 1999 Transition from dripping to jetting.
  {\em J. Fluid Mech.\/} {\bf 383}, 307--326.

\bibitem[Couairon \& Chomaz(1999)]{Couairon1999}
{\sc Couairon, A. \& Chomaz, J.-M.} 1999 Fully nonlinear global modes in slowly
  varying flows. {\em Phys. Fluids\/} {\bf 11}, 3688--3703.

\bibitem[Deissler(1987)]{Deissler1987}
{\sc Deissler, R.~J.} 1987 The convective nature of instability in plane
  {P}oiseuille flow. {\em Phys. Fluids\/} {\bf 30}~(8), 2303--2305.

\bibitem[Duda \& Vrentas(1967)]{DudayVrentas}
{\sc Duda, J.~L. \& Vrentas, J.~S.} 1967 Fluid mechanics of laminar liquid
  jets. {\em Chem. Eng. Sci.\/} {\bf 22}, 855--869.

\bibitem[Eggers(1997)]{Eggers1997}
{\sc Eggers, J.} 1997 Nonlinear dynamics and breakup of free surface flows.
  {\em Rev. Mod. Phys.\/} {\bf 69}, 865--929.

\bibitem[Eggers \& Villermaux(2008)]{EyV}
{\sc Eggers, J. \& Villermaux, E.} 2008 Physics of liquid jets. {\em Rep. Prog.
  Phys.\/} {\bf 71}, 036601.

\bibitem[Gavis(1964)]{Gavis1964}
{\sc Gavis, J.} 1964 Contribution of surface tension to expansion and
  contraction of capillary jets. {\em Phys. Fluids\/} {\bf 7}~(7), 1097--1098.

\bibitem[Gordillo \& Gekle(2010)]{GordilloTipBreakup}
{\sc Gordillo, J.~M. \& Gekle, S.} 2010 Generation and breakup of {W}orthington
  jets after cavity collapse. {P}art 2: {T}ip breakup of stretched jets. {\em
  J. Fluid Mech.\/} {\bf 663}, 331--346.

\bibitem[Gordillo \& P\'erez-Saborid(2005)]{GordilloFirstWind}
{\sc Gordillo, J.~M. \& P\'erez-Saborid, M.} 2005 Aerodynamic effects in the
  break-up of liquid jets: on the first wind-induced break-up regime. {\em J.
  Fluid Mech.\/} {\bf 541}~(541), 1--20.

\bibitem[Gordillo {\em et~al.\/}(2001)Gordillo, P\'erez-Saborid \&
  Gan\'an-Calvo]{Gordillo2001}
{\sc Gordillo, J.~M., P\'erez-Saborid, M. \& Gan\'an-Calvo, A.~M.} 2001 Linear
  stability of co-flowing liquid--gas jets. {\em J. Fluid Mech.\/} {\bf 448},
  23--51.

\bibitem[Goren(1966)]{Goren66}
{\sc Goren, S.~L.} 1966 Development of the boundary layer at a free surface
  from a uniform shear flow. {\em J. Fluid Mech.\/} {\bf 25}, 87--95.

\bibitem[Goren \& Wronski(1966)]{GorenyWronski}
{\sc Goren, S.~L. \& Wronski, S.} 1966 The shape of low-speed capillary jets of
  {N}ewtonian liquids. {\em J. Fluid Mech.\/} {\bf 25}, 185--198.

\bibitem[Harmon(1955)]{Harmon1955}
{\sc Harmon, D.~B.} 1955 Drop sizes from low speed jets. {\em J. Franklin
  Inst.\/} {\bf 259}~(6), 519 -- 522.

\bibitem[Herbert(1997)]{Herbert1997}
{\sc Herbert, T.} 1997 Parabolized stability equations. {\em Annu. Rev. Fluid
  Mech.\/} {\bf 29}, 245--283.

\bibitem[Herrada {\em et~al.\/}(2008)Herrada, Del~Pino,  \&
  Fern\'andez-Feria]{Herrada2008}
{\sc Herrada, M.~A., Del~Pino, C.,  \& Fern\'andez-Feria, R.} 2008 Stability of
  the boundary layer flow on a long thin rotating cylinder. {\em Phys.
  Fluids\/} {\bf 20}, 034105.

\bibitem[Huerre \& Monkewitz(1990)]{Huerre1990}
{\sc Huerre, P. \& Monkewitz, P.} 1990 Local and global instabilities in
  spatially developing flows. {\em Annu. Rev. Fluid Mech.\/} {\bf 22},
  473--537.

\bibitem[Le~Diz\`es(1997)]{ledizes97}
{\sc Le~Diz\`es, S.} 1997 Global modes in falling capillary jets. {\em Eur. J.
  Mech. B/Fluids\/} {\bf 16}, 761--778.

\bibitem[Leib \& Goldstein(1986{\natexlab{{\em a\/}}})]{LeibyGoldsteinAC}
{\sc Leib, S.~J. \& Goldstein, M.~E.} 1986{\natexlab{{\em a\/}}} Convective and
  absolute instability of a viscous liquid jet. {\em Phys. Fluids\/} {\bf
  29}~(4), 952--954.

\bibitem[Leib \& Goldstein(1986{\natexlab{{\em b\/}}})]{LeibyGoldstein}
{\sc Leib, S.~J. \& Goldstein, M.~E.} 1986{\natexlab{{\em b\/}}} The generation
  of capillary instabilities on a liquid jet. {\em J. Fluid Mech.\/} {\bf 168},
  479--500.

\bibitem[Lesshafft {\em et~al.\/}(2006)Lesshafft, Huerre, Sagaut \&
  Terracol]{Lesshafft2006}
{\sc Lesshafft, L., Huerre, P., Sagaut, P. \& Terracol, M.} 2006 Nonlinear
  global modes in hot jets. {\em J. Fluid Mech.\/} {\bf 554}, 393--409.

\bibitem[Lin \& Reitz(1998)]{Lin1998}
{\sc Lin, S.~P. \& Reitz, R.~D.} 1998 Drop and spray formation from a liquid
  jet. {\em Annu. Rev. Fluid Mech.\/} {\bf 30}, 85--105.

\bibitem[Miles(1960)]{Miles1960}
{\sc Miles, J.~W.} 1960 The hydrodynamic stability of a thin film of liquid in
  uniform shearing motion. {\em J. Fluid Mech.\/} {\bf 8}, 593--610.

\bibitem[Monkewitz {\em et~al.\/}(1988)Monkewitz, Davis, Bojorquez \&
  Yu]{Monkewitz88cap}
{\sc Monkewitz, P.~A., Davis, J., Bojorquez, B. \& Yu, M.-H.} 1988 The breakup
  of a liquid jet at high {W}eber number. {\em Bull. Am. Phys. Soc.\/} {\bf
  33}, 2273.

\bibitem[Monkewitz {\em et~al.\/}(1993)Monkewitz, Huerre \& Chomaz]{monk93}
{\sc Monkewitz, P.~A., Huerre, P. \& Chomaz, J.-M.} 1993 Global linear
  stability analysis of weakly non-parallel shear flows. {\em J. Fluid Mech.\/}
  {\bf 251}, 1--20.

\bibitem[O\~guz(1998)]{OguzRelaxation}
{\sc O\~guz, H.} 1998 On the relaxation of laminar jets at high {R}eynolds
  numbers. {\em Phys. Fluids\/} {\bf 10}~(2), 361--367.

\bibitem[O\~guz \& Prosperetti(1993)]{Oguz1993}
{\sc O\~guz, H. \& Prosperetti, A.} 1993 Dynamics of bubble growth and
  detachment from a needle. {\em J. Fluid Mech.\/} {\bf 257}, 111--145.

\bibitem[O'Donnell {\em et~al.\/}(2001)O'Donnell, Chen \& Lin]{Odonnell2001}
{\sc O'Donnell, B., Chen, J.~N. \& Lin, S.~P.} 2001 Transition from convective
  to absolute instability in a liquid jet. {\em Phys. Fluids\/} {\bf 13}~(9),
  2732--2734.

\bibitem[Osborne \& Steinberg(2006)]{Osborne2006}
{\sc Osborne, B.~P. \& Steinberg, T.~A.} 2006 An experimental investigation
  into liquid jetting modes and break-up mechanisms conducted with a new
  reduced gravity facility. {\em Microgravity Sci. Technol.\/} {\bf
  XVIII}~(3/4), 57--61.

\bibitem[Philippe \& Dumargue(1991)]{franceses}
{\sc Philippe, C. \& Dumargue, P.} 1991 Etude de l'etablissement d'un jet
  liquide laminaire emergeant d'une conduite cylindrique verticale semi-infinie
  et soumis a l'influence de la gravite. {\em ZAMP\/} {\bf 42}, 227--242.

\bibitem[Pier \& Huerre(2001)]{Pier2001}
{\sc Pier, B. \& Huerre, P.} 2001 Nonlinear self-sustained structures and
  fronts in spatially developing wake flows. {\em J. Fluid Mech.\/} {\bf 145},
  145--174.

\bibitem[Pier {\em et~al.\/}(1998)Pier, Huerre, Chomaz \& Couairon]{Pier98}
{\sc Pier, B., Huerre, P., Chomaz, J.-M. \& Couairon, A.} 1998 Steep nonlinear
  global modes in spatially developing media. {\em Phys. Fluids\/} {\bf
  10}~(10), 2433--2435.

\bibitem[Rayleigh(1878)]{Rayleigh1878}
{\sc Rayleigh, W.~S.} 1878 On the instability of jets. {\em Proc. London Math.
  Soc.\/} {\bf 10}, 4--13.

\bibitem[Sauter \& Buggisch(2005)]{SauteryBuggisch}
{\sc Sauter, U.~S. \& Buggisch, H.~W.} 2005 Stability of initially slow viscous
  jets driven by gravity. {\em J. Fluid Mech.\/} {\bf 533}, 237--257.

\bibitem[Sevilla {\em et~al.\/}(2005)Sevilla, Gordillo \&
  Mart\'{\i}nez-Baz\'an]{Sevilla2005a}
{\sc Sevilla, A., Gordillo, J.~M. \& Mart\'{\i}nez-Baz\'an, C.} 2005 Transition
  from bubbling to jetting in a coaxial air-water jet. {\em Phys. Fluids\/}
  {\bf 17}, 018105.

\bibitem[Sevilla \& Mart\'inez-Baz\'an(2004)]{Sevilla2004}
{\sc Sevilla, A. \& Mart\'inez-Baz\'an, C.} 2004 Vortex shedding in high
  {R}eynolds number axisymmetric bluff-body wakes: Local linear instability and
  global bleed control. {\em Phys. Fluids\/} {\bf 16}, 3460.

\bibitem[Smith \& Davis(1982)]{SmithDavis1982}
{\sc Smith, M.~K. \& Davis, S.~H.} 1982 The instability of sheared liquid
  layers. {\em J. Fluid Mech.\/} {\bf 121}, 187--206.

\bibitem[Smith \& Moss(1917)]{SmithMoss}
{\sc Smith, S. W.~J. \& Moss, H.} 1917 Experiments with mercury jets. {\em
  Proc. Roy. Soc. A\/} {\bf 93}, 373--393.

\bibitem[S\"oderberg(2003)]{Soderberg03}
{\sc S\"oderberg, L.~D.} 2003 Absolute and convective instability of a
  relaxational plane liquid jet. {\em J. Fluid Mech.\/} {\bf 493}, 89--119.

\bibitem[Theofilis(2011)]{Theofilis2011}
{\sc Theofilis, V.} 2011 Global linear instability. {\em Annu. Rev. Fluid
  Mech.\/} {\bf 43}, 319--352.

\bibitem[Tillett(1968)]{Tillett68}
{\sc Tillett, J. P.~K.} 1968 On the laminar flow in a free jet of liquid at
  high {R}eynolds number. {\em J. Fluid Mech.\/} {\bf 32}, 273--292.

\bibitem[Utada {\em et~al.\/}(2007)Utada, Chu, Fernandez-Nieves, Link, Holtze
  \& Weitz]{UtadaMagic}
{\sc Utada, A.~S., Chu, L.-Y., Fernandez-Nieves, A., Link, D.~R., Holtze, C. \&
  Weitz, D.~A.} 2007 Dripping, jetting, drops, and wetting: {T}he magic of
  microfluidics. {\em MRS Bull.\/} {\bf 32}, 702--708.

\bibitem[Vihinen {\em et~al.\/}(1997)Vihinen, Honohan \& Lin]{Vihinen1997}
{\sc Vihinen, I., Honohan, A.~M. \& Lin, S.~P.} 1997 Image of absolute
  instability in a liquid jet. {\em Phys. Fluids\/} {\bf 9}~(11), 3117--3119.

\bibitem[Weber(1931)]{Weber1931}
{\sc Weber, C.} 1931 On the breakdown of a fluid jet. {\em Z. Angew. Math.
  Mech.\/} {\bf 11}, 136--141.

\bibitem[Wilkes {\em et~al.\/}(1999)Wilkes, Phillips \& Basaran]{Wilkes1999}
{\sc Wilkes, E.~D., Phillips, S.~D. \& Basaran, O.~A.} 1999 Computational and
  experimental analysis of dynamics of drop formation. {\em Phys. Fluids\/}
  {\bf 11}~(12), 3577--3598.

\end{thebibliography}
\end{document}